\documentclass[a4paper]{scrartcl}

\usepackage{lmodern}
\usepackage{amssymb,amsmath}
\usepackage[T1]{fontenc}
\usepackage[utf8]{inputenc}
\usepackage{upquote}
\usepackage[tracking=smallcaps,letterspace=70]{microtype}
\usepackage[unicode=true]{hyperref}
\usepackage{longtable,booktabs}
\usepackage{graphicx,grffile}
\usepackage{authblk}
\usepackage[comma,super,sort&compress]{natbib}
\usepackage{tabularx}
\newcolumntype{Y}{>{\centering\arraybackslash}X}
\usepackage[font=small]{caption}
\usepackage{multirow}
\usepackage{subfig}
\usepackage{comment}

\usepackage[british]{babel}
\usepackage[modulo]{lineno}

\UseMicrotypeSet[protrusion]{basicmath}

\providecommand{\subtitle}[1]{}

\PassOptionsToPackage{usenames,dvipsnames}{color}

\hypersetup{%
	pdftitle={Magnetorheology of magnetic-particle-seeded blood analogues},
	pdfauthor={R. Rodrigues; F.J. Galindo-Rosales; L. Campo-Dea{\~n}o},
	pdfkeywords={Keyword \#1, keyword \#2, keyword \#3},
	pdfborder={0 0 0},
	colorlinks=true,
	linkcolor=Blue,
	citecolor=Blue,
	urlcolor=Blue,
	breaklinks=true
}%

\makeatletter
	\def\maxwidth{\ifdim\Gin@nat@width>\linewidth\linewidth\else\Gin@nat@width\fi}
	\def\maxheight{\ifdim\Gin@nat@height>\textheight\textheight\else\Gin@nat@height\fi}
\makeatother

\setkeys{Gin}{width=\maxwidth,height=\maxheight,keepaspectratio}

\ifx\paragraph\undefined\else
	\let\oldparagraph\paragraph
	\renewcommand{\paragraph}[1]{\oldparagraph{#1}\mbox{}}
\fi
\ifx\subparagraph\undefined\else
	\let\oldsubparagraph\subparagraph
	\renewcommand{\subparagraph}[1]{\oldsubparagraph{#1}\mbox{}}
\fi

\title{Magnetorheological Characterization of Blood Analogues Seeded with Paramagnetic Particles}

\author[1,2]{R. Rodrigues}
\author[2,3]{F.J. Galindo-Rosales}
\author[1,2$\dagger$]{L. Campo-Dea{\~n}o}

\affil[1]{CEFT - Centro de Estudos de Fen{\'o}menos de Transporte, Depto.~de Engenharia Mec{\^a}nica, Faculdade de Engenharia, Universidade do Porto, Rua Dr.\ Roberto Frias, 4200-465, Porto, Portugal} 
\affil[2]{ALiCE - Laborat{\'o}rio Associado em Engenharia Qu{\'i}mica, Faculdade de Engenharia, Universidade do Porto, Rua Dr.\ Roberto Frias, 4200-465, Porto, Portugal}
\affil[3]{CEFT - Centro de Estudos de Fen{\'o}menos de Transporte, Depto.~de Engenharia Química  e Biol\'ogica, Faculdade de Engenharia, Universidade do Porto, Rua Dr.\ Roberto Frias, 4200-465, Porto, Portugal} 
\affil[$\dagger$]{Email: \href{mailto:campo@fe.up.pt}{campo@fe.up.pt}}

\date{}

\begin{document}

\maketitle

\begin{abstract} 
\small
	\noindent Abstract - Magnetic particle under external fields can be useful in various medical applications, gaining access to the whole body if deployed in the bloodstream. Localised drug delivery, haemorrhage control, and cancer treatment are among the applications that have the potential to become revolutionary therapies. Despite this interest, a magnetorheological characterisation of particle-seeded blood has yet to be achieved. In this work, we evaluate the magnetorheological response of blood analogues seeded with paramagnetic particles in different concentrations, under the effects of a uniform, density-varying magnetic field. Through steady shear experiments, we encounter the usual magnetically-induced shear thinning response, and oscillatory shear results point toward significant alterations in the fluids' microstructure. However, experimental limitations make it difficult to accurately evaluate the oscillatory shear response of such rheologically subtle fluids, limiting both the quality and quantity of achievable information. Despite experimental limitations, our results demonstrate that magnetic fields can induce marked and quantifiable rheological changes in seeded blood analogues. The framework established here provides a foundation for future studies on real blood samples and for the design of magnetically responsive biomedical systems. \vspace{.5cm}
 
\end{abstract}

\newpage

\section{Introduction}
\label{sec:intro}

It has long been known that blood has magnetic properties\citep{pauling1936}. Red blood cells (RBCs or erythrocytes) are one the major blood constituents and are the main responsible for its mechanical properties\citep{robertson2008}. RBCs contain haemoglobin, a protein whose iron content makes it paramagnetic when not carrying oxygen (deoxyhaemoglobin)\citep{pauling1936,zborowski2003}. When under a magnetic field, dipole interactions arise between RBCs, leading them to aggregate into chains aligned with the field. This can be used to knowingly alter blood's rheological properties without resourcing to drugs or other possibly harmful means\citep{tao2011}, which is extremely attractive as blood rheology is connected to a series of paramount conditions and diseases. Heart attack, stroke, aneurysms, hypertension and other, possibly crippling or even fatal conditions are associated with high blood viscosity and other rheological characteristics\citep{dintenfass1963,campo2015}. On the other hand, the increase of viscosity may hinder blood flow to aid in controlling haemorrhages in complex injuries or surgery\citep{misra2009}. 

Blood is, however, a very complex fluid and the effects of magnetic field influence on its microstructure and behaviour are still under investigation. The effects on the viscoelastic properties of blood are relatively undiscussed as most experimental works focus on viscosity, which still return conflicting results. \citeauthor{tao2011} report that a short magnetic field pulse along the flow direction (1.3 T for approximately 1 min) induced a significant viscosity decrease (20-30\%), lasting a few hours before returning to the original value. \citeauthor{yan2024} also observed a significant viscosity decrease with a capillary viscometer under an alternating magnetic field. \citeauthor{tao2011} attribute the viscosity decrease to magnetic-induced RBC aggregation into larger, streamlined structures and increased polydispersity, while \citeauthor{yan2024} additionally contemplate the magnetically-enhanced rheological properties of the RBCs membranes, which facilitates inter-cell sliding that further decreases resistance to flow. On the other hand, \citeauthor{haik2001} observed an increase in viscosity of venous blood (30\%) when subject to a strong magnetic field parallel to the flow (10 T), ascribing it to magnetic torque maintaining an angular velocity difference between the RBCs and the plasma, leading to an additional viscous dissipation. The contradicting literature dictates that more research needs to be conducted and, particularly, the details of the experimental setup and properties of the blood samples need to be thoroughly described as the magnetorheological response can be heavily influenced by flow dimensions, magnetic field properties, temperature, RBC concentration and blood oxygenation\citep{yamamoto2004,yan2024,fahraeus1931}.

In addition to the direct alteration of haemorheological properties, magnetic fields can be used in a multitude of medical applications\citep{andra2007}. Among these, incorporating magnetic particles into blood flow holds significant promise for advanced medical treatments, such as targeted drug delivery, haemorrhage control and cancer treatment through hyperthermia therapies or localised blood/oxygen starvation\citep{liu2001,arruebo2007,tekleab2019,andra2007}. However, despite these potential benefits, a comprehensive magnetorheological characterization of blood loaded with magnetic particles has yet to be achieved. This study aims to address this gap by thoroughly analysing the behaviour of both Newtonian and viscoelastic blood analogues under magnetic fields, while also assessing the influence of factors such as particle type and concentration.

\subsection{Magnetorheology}

Magnetorheological (MR) fluids are usually suspensions of magnetisable particles in a non-magnetisable continuous phase. Under a magnetic field, dipole interactions between the particles will force them to aggregate into elongated chains aligned with the field lines, and the induced microstructure leads to significant alterations of the rheological properties\citep{ginder1998}. Typically, magnetised MR fluids present a solid-like behaviour at rest, where a stress threshold must be surpassed to induce fluid flow (yield stress). Considering steady shear, a strong shear-thinning response is usually observed as the shear rate is increased due to the enhanced viscous shearing forces breaking down the microstructure. This competition between viscous and magnetic effects has been widely characterised by the Mason number, $Mn$, and the response to steady shear can be reasonably described by the Casson model, which has been found to better fit experimental data than the Bingham model, particularly through a smoother transition from the yielding to the Newtonian response\citep{berli2012,morillas2019}.

Regarding the response to oscillatory shear, MR fluids typically exhibit a small linear viscoelastic regime (LVE) with a predominantly elastic response (storage modulus larger than the loss modulus, $G'>G''$). Increasing the strain typically leads to a steady decrease of $G'$, while $G''$ reaches a maximum before decreasing more slowly than $G'$\citep{li2004,sim2003,vicente2011review,parthasarathy1999}. As such, MR fluids usually display a type III behaviour\citep{hyun2002}, where the overshoot in $G''$ at small strain has been attributed to small rearrangements of the unstable microstructure\citep{parthasarathy1999,li2004}, dissipating energy into the continuous phase and increasing $G''$. Further increasing the strain leads to a decrease of $G''$ as significant microstructural changes occur within the oscillatory cycle, particularly the breakdown of the microstructure at large instantaneous strain rates. At small oscillation frequencies, the response is practically frequency-independent, and a set of rheological transitions can be defined with strain increase. At sufficiently small deformations, the response is linearly viscoelastic, and increasing the strain onsets a non-linear viscoelastic behaviour from the slight rearrangements of the microstructure. Further increasing the strain results in large microstructural changes within the oscillatory cycle, which results in a deeply non-linear response until, at sufficiently large deformation, the elastic contribution is very small, and the response becomes viscoplastic. At larger frequencies, a Newtonian response is expected with increasing strain from negligible microstructure re-formation until, at a sufficiently large frequency, a Newtonian response is found independently of the strain\citep{li2004,parthasarathy1999}.

\subsection{FT-Chebyshev analysis}

The non-linear viscoelastic (NLVE) behaviour of MR fluids has been extensively discussed, but the intricacies of the non-linear response remain obscure in measures of average viscoelasticity ($G'$ and $G''$). To uncover the complexities of non-linear viscoelasticity, (Fourier transform) FT-Chebyshev analysis comes into play. For the sake of clarity, we will not discuss the mathematical foundation of FT-Chebyshev formulation; for an in-depth description, we refer to the work of \citeauthor{ewoldt2008}. Nevertheless, a few concepts are worth mentioning here. In a standard oscillatory shear test, a sinusoidal strain is imposed on the sample, $\gamma(t)=\gamma_0\,\sin(\omega\,t)$, where $\gamma_0$ is the strain amplitude and $\omega$ the oscillation frequency. The strain rate is then given by the time derivative, $\dot\gamma(t)=\dot\gamma_0\,\cos(\omega\,t)$, orthogonal to the strain, where $\dot\gamma_0=\gamma_0\,\omega$ is the strain rate amplitude. In a small amplitude oscillatory shear test (SAOS) the measurement is conducted within the LVE where the stress response is linear, which can fully described by the first-harmonic storage and loss moduli ($G'_1$ and $G''_1$)\footnote{The viscoelastic moduli until now represented by $G'$ and $G''$ referred to these first-harmonic moduli.}. However, increasing the strain into the large amplitude oscillatory shear regime (LAOS) leads to a non-linear behaviour where additional harmonics distort the stress response, which is no longer a simple sinusoid. The stress can then be written as a Fourier series of multiple harmonic contributions, which can be used to identify non-linear behaviour. However, this methodology lacks the physical meaning of the higher-order contributions. At the same time, the resulting first-harmonic moduli ($G'_1$ and $G''_1$) are inadequate in the non-linear regime and give only a measure of the average viscoelastic behaviour\citep{ewoldt2008}. 

Another way of probing non-linearity is through the elastic and viscous Lissajous (or Lissajous–Bowditch) curves, in which the raw stress is plotted as a function of the instantaneous strain and strain rate, respectively. A Hookean solid would return a positive diagonal elastic Lissajous curve and a perfectly circular viscous Lissajous curve, whereas a purely viscous material would yield the inverse. A linear viscoelastic response is then given by perfect ellipses in both elastic and viscous Lissajous, and distortions of the ellipse shape characterise a non-linear response, as higher harmonics affect the stress signal. \citeauthor{cho2005} proposed decomposing the raw stress in LAOS tests into elastic and viscous components, which effectively collapse the Lissajous into one-dimensional curves, facilitating the analysis. Additionally, the Lissajous curves can also provide quantitative information. \citeauthor{ewoldt2008} defined the minimum strain modulus as the tangent modulus at minimum strain, $G'_M=d\sigma/d\gamma\vert_{\gamma=0}$, and the large strain modulus as the secant modulus at maximum strain, $G'_L=\sigma/\gamma\vert_{\gamma=\gamma_0}$. Similarly, minimum and large rate dynamic viscosities can be defined as $\eta'_M=d\sigma/d\dot\gamma\vert_{\dot\gamma=0}$ and $\eta'_L=\sigma/\dot\gamma\vert_{\dot\gamma=\dot\gamma_0}$, respectively. As such, in the non-linear regime, these measures help quantify the (intracycle) non-linear behaviour, while in the linear regime, they reduce to $G'_1$ and $\eta'_1=G''_1/\omega$, respectively.

\citeauthor{ewoldt2008} also proposed to fit Chebyshev polynomials of the first kind to the decoupled stresses to allow an analysis of the harmonic contributions independently of the number of selected harmonics, which is usually arbitrary. The first order elastic and viscous Chebyshev coefficients are related to the widely used first-harmonic moduli: $e_1=G'_1$ and $v_1=\eta'_1=G''_1/\omega$, respectively, which are the sole contributors in the LVE; and the non-linear response can be described through the third order coefficients. A positive third-order elastic Chebyshev coefficient, $e_3>0$, relates to a higher elastic stress at maximum strain than the linear prediction (considering $e_1$ alone), whereas a negative value, $e_3<0$, corresponds to a lesser elastic stress at maximum strain than the linear representation. The same can be gathered from the third-order viscous coefficient regarding the viscous stress to the strain rate. This has led $e_3>0$ and $e_3<0$ to be respectively associated with intracycle elastic strain stiffening and softening, and $v_3>0$ and $v_3<0$, to intracycle viscous shear thickening and thinning, which often contradict the average viscoelastic behaviour described by $G'_1$ and $G''_1$. This issue has been addressed before\citep{mermet2015,ewoldt2013} and an alternative framework has been suggested that allows the interpretation of the third order coefficients within the context of the first-harmonic non-linear behaviour\citep{ewoldt2013}. Because the strain and strain rate are orthogonal by definition, the sign of the third order coefficients may be understood, not as a description of the non-linear response itself, but as an indicator of the deformation input that leads to the average non-linearity. For example, a positive $e_3$ relates to a relatively larger elastic stress at maximum strain ($\gamma=\gamma_0$) or minimum rate ($\dot\gamma=0$); as such, if the material presents elastic softening (that is, a decrease of $G'_1$), $e_3>0$ indicates that the elastic softening is driven by the strain rate. The same can be said of the viscous response; a positive $v_3$ with an average viscous thinning material ($G''_1$ decrease) relates to the thinning non-linearity being driven by the strain, as the viscous stress is relatively lesser at $\gamma=\gamma_0$.

It is worth mentioning that while on the one hand, the Chebyshev coefficients give no direct information on the magnitude of the response's non-linearity, contrarily to the viscoelastic measures ($G'_M, G'_L, \eta'_M, \eta'_L$), the significance of the Chebyshev analysis lies in the fact that the harmonic contributions are attained independently and, as such, are not as sensitive to high-order excitations that may arise from experimental error. This may lead to discrepancies between the analysis of Chebyshev coefficients and the non-linear viscoelastic moduli if the latter are calculated from the raw stress signal or from very high order data, where local effects and experimental errors may gain relevance. On the other hand, an analysis from the third order coefficients alone may be insufficient to fully describe highly-non-linear responses that incorporate additional significant high-order harmonics. As such, the Chebyshev coefficients and the viscoelastic moduli are complementary information sources.

FT-Chebyshev analysis is a powerful tool to unveil and quantify the intricacies of a material's non-linear response. However, there is still a lack of in-depth analysis of non-linear magnetorheological data in the literature. As far as we know, the work of \citeauthor{wang2021} is one of the few studies that probe experimental magnetorheological data through FT-Chebyshev analysis. The authors performed LAOS tests with a magnetorheological grease (a suspension of carbonyl iron particles in a semi-solid grease) under different magnetic field intensities. The authors found the application of the magnetic field transformed the rheological response from type I to type III\citep{hyun2002}, and the LVE was significantly shortened, widening with magnetic field intensity increase but still narrower than for the non-magnetised grease. In the absence of magnetic influence, the MR grease is already highly viscoelastic, displaying elastic softening and viscous thinning through the decrease of $G'_1$ and $\eta'_1$, both driven by the strain rate ($G'_M<G'_L$ and $e_3>0$, $\eta'_L<\eta'_M$ and $v_3<0$). The magnetic field application led to significant alterations in the non-linear response. The authors report a sign shift of $e_3$ to negative values at larger strain (for intermediate fields) and positive $v_3$ at small strain, later shifting to $v_3<0$. The authors discuss the possibility of complex response arising from interactions between the fibre structure of the continuous phase and the magnetically-induced particle-chain structure.

\section{Materials and methods}
\label{sec:methods}

\subsection{Experimental setup}

In this work we employed an Anton Paar MCR302-e rotational rheometer (minimum torque $T_\mathrm{min} = 1$ and 0.5 nN$\cdot$m for steady and oscillatory shear, respectively, and maximum rotational velocity $\Omega_\mathrm{max}=314$ rad/s), equipped with a magnetorheological cell able to generate magnetic fields of density up to $B \leq 760$ mT, uniform and perpendicular to planar geometries. The magnetic circuit was closed via a yoke placed on top of the geometry\citep{laeuger2005}, and the effective field density was measured with a teslameter (FH 54, MAGNET-PHYSIK), inserted in a slot directly below the static bottom plate. The magnetorheological cell does not allow for precise temperature control. Consequently, these measurements were performed at room temperature. Despite lacking temperature fine-tuning, an external cooling system (AWC100, Julabo), circulating cold water, was used to dissipate the heat from the magnetic field generation. During the whole experimental campaign, the temperature averaged 22.7$^\circ$C, with maximum deviation $[-2.6,1.5]^\circ$C.

The employed measuring geometry was a PP20 MRD P2, a serrated Parallel-Plate designed for magnetic testing, with $20$ mm of diameter and moment of inertia $I\approx0.0006$ mN$\cdot$m$\cdot$s\textsuperscript{2}. The gap was set to $h_\mathrm{c}=0.1$ mm.

\subsection{Working fluids}

Two blood analogues were employed, one Newtonian and one viscoelastic\citep{campo2013}. The Newtonian analogue, NBa, was an aqueous solution of 52 wt\% Dimethyl sulfoxide (DMSO) and the viscoelastic analogue, VBa, was fabricated by adding 100 ppm of Xanthan Gum to the Newtonian one. The density of both solutions was approximately the same, $\rho \approx 1050$ kg/m\textsuperscript{3}, at 22.7$^\circ$C. Regarding the VBa's elastic characteristics, we used a Capillary Breakup Extensional Rheometer (CaBER) coupled with a high-speed camera (Photron FASTCAM Mini UX100) with a set of lenses (Optem Zoom 70 X), to obtain images of the sample breakup under one-dimensional extensional flow, using the slow retraction method\citep{campo2010slow}. The images corresponding to the elasto-capillary regime were treated to obtain the evolution of the filament's diameter with time and, through an exponential fit\citep{mckinley2005}, the fluid's longest relaxation time was obtained: $\lambda_\mathrm{VBa} = 0.350 \pm 0.044$ ms, which is correspondent to a subtle elasticity (data and setup not shown here for brevity).

\subsection{Magnetic particles}

Two magnetic particles were employed, the M270 and MyOne Carboxylic Acid Dynabeads (Thermo Fisher Scientific). The M270, the larger of the two, had a diameter of $d_\mathrm{p}=2.8$ \textmu m while the MyOne only 1 \textmu m. \cite{grob2018} present magnetisation curves for both particles (saturation magnetization values of $M_\mathrm{s}=9.4$ and 6.7 Am\textsuperscript{2}/kg for the MyOne and M270, respectively, for $B\gtrsim 0.5$ T) The particles were seeded in either fluid, being re-dispersed prior to all measurements to counteract sedimentation and break up any previously established structure. The M270 and MyOne particles had densities of 1600 and 1800 kg/m\textsuperscript{3}, respectively.

\subsection{Experimental procedure}

The sample was carefully loaded onto the geometry's bottom plate with a VWR\textsuperscript{\textregistered{}} standard line precision pipette. The steady shear measurements ranged shear rates $1 \leq \dot\gamma \leq 10000$ s\textsuperscript{-1}, with a constant time interval of 20 s between shear rate shift and torque measurement. The oscillatory shear response was evaluated by performing amplitude sweeps between $0.1\leq \gamma_0 \leq 100$ at frequencies between $0.1 \leq \omega \leq 2$ rad/s. 

The magnetic particles are a precious resource, and the number of measurements was minimised. For steady shear, a minimum of two repetitions was deemed acceptable, having that both measurements agreed reasonably well; if significant data discrepancies were found, additional measurements were conducted up to a maximum of five.

A compact experimental campaign allowed us to investigate the effects of the magnetic field density, particle concentration, particle type and continuous phase (NBa or VBa). We selected benchmark measurements that allowed for the independent variation of relevant characteristics. The samples consisted of 15 wt\% of MyOne particles subjected to a magnetic field density of 250 mT. We focused particularly on testing the continuous phase, specifically the blood analogues, while varying the other properties for both NBa and VBa samples. To test for the influence of the field density, it was reduced from the benchmark 250 mT to 50 mT. Similarly, testing the particle concentration, it was reduced from 15 wt\% to 5 wt\% (volume fractions of approximately: 10.6 and 3.4\% for the M270, and 9.7 and 3.0\% for the MyOne particles). Lastly, we tested the M270 particles instead of the MyOne particles. Table 1 summarizes the nomenclature and the description of the measurements in relation to the seeded particle concentration, type and applied magnetic field density for both blood analogues as the continuous phase.

\begin{table}[htp]
    \centering
    \renewcommand{\arraystretch}{1.5}
    \small{
    \caption{Nomenclature and description of the measurements in relation to the seeded particle concentration, type and applied magnetic field density, valid for both blood analogues as the continuous phase (Newtonian, NBa, and viscoelastic, VBa)}
    \label{tab:working_fluids}
    \begin{tabularx}{\textwidth}{>{\hsize=1\hsize}Y|>{\hsize=.9\hsize}Y>{\hsize=.8\hsize}Y>{\hsize=.9\hsize}Y|>{\hsize=1.4\hsize}Y}
    \hline
    Nomenclature&Particle concentration&Particle type&Magnetic field density&Interest\\
    \hline
    \hline
    Unseeded&0 wt\%&-&0 mT&Analogue rheology\\
    15\_MyOne\_250&15 wt\%&MyOne&250 mT&Benchmark\\
    5\_MyOne\_250&5 wt\%&MyOne&250 mT&Particle concentration\\
    15\_MyOne\_50&15 wt\%&MyOne&50 mT&Magnetic field density\\
    15\_M270\_250&15 wt\%&M270&250 mT&Particle type\\
    \hline
    \end{tabularx}}
\end{table}

\section{Results and discussion}
\label{sec:results}

Initially, we employed a smooth Cone-and-Plate geometry for our measurements (CP20 MRD). However, this geometry showed clear signs of apparent slip, a well-known phenomenon associated with suspension rheology, where a particle-depletion layer forms near the geometry's smooth walls and in which a large part of the imposed deformation is applied solely to the low-viscosity continuous phase\citep{barnes1995,vicente2004}. The serrated PP20 MRD P2 geometry seemingly rectified the apparent slip, but our initial testing also showed that this geometry is affected by a gap-error of approximately $\varepsilon=123$ \textmu m. Such errors are expected of detailed geometries because the sample flows within the geometry grooves, similar to a porous medium\citep{carotenuto2013,nickerson2005}. As such, all data gathered with the PP20 MRD P2 are corrected for this gap-error according to the formulation of \citeauthor{kramer1987}, similarly to our previous work\citep{rodrigues2025}. Moreover, we also gathered that low and high shear limitations constrain our experimental window from this initial testing. At low shear, the rheometer's minimum torque limit\citep{ewoldt2015} is outside our experimental range. Still, surface tension forces acting along the sample's contact line\citep{johnston2013} are responsible for uncertainties that can be estimated by a minimum torque 70 times larger than the instrument's specification ($70\times \mathrm{T}_\mathrm{min}$). At high shear, sample inertia may lead to an apparent shear-thickening\citep{ewoldt2015} which is reasonably predicted by the secondary flow limit\citep{turian1972}. For conciseness, this preliminary data is not shown here, but a brief discussion is presented as supplementary material (Section \ref{sec:sup:preliminary}).

\subsection{Steady shear}

\begin{figure}[t]
\centering
\includegraphics[width=\linewidth]{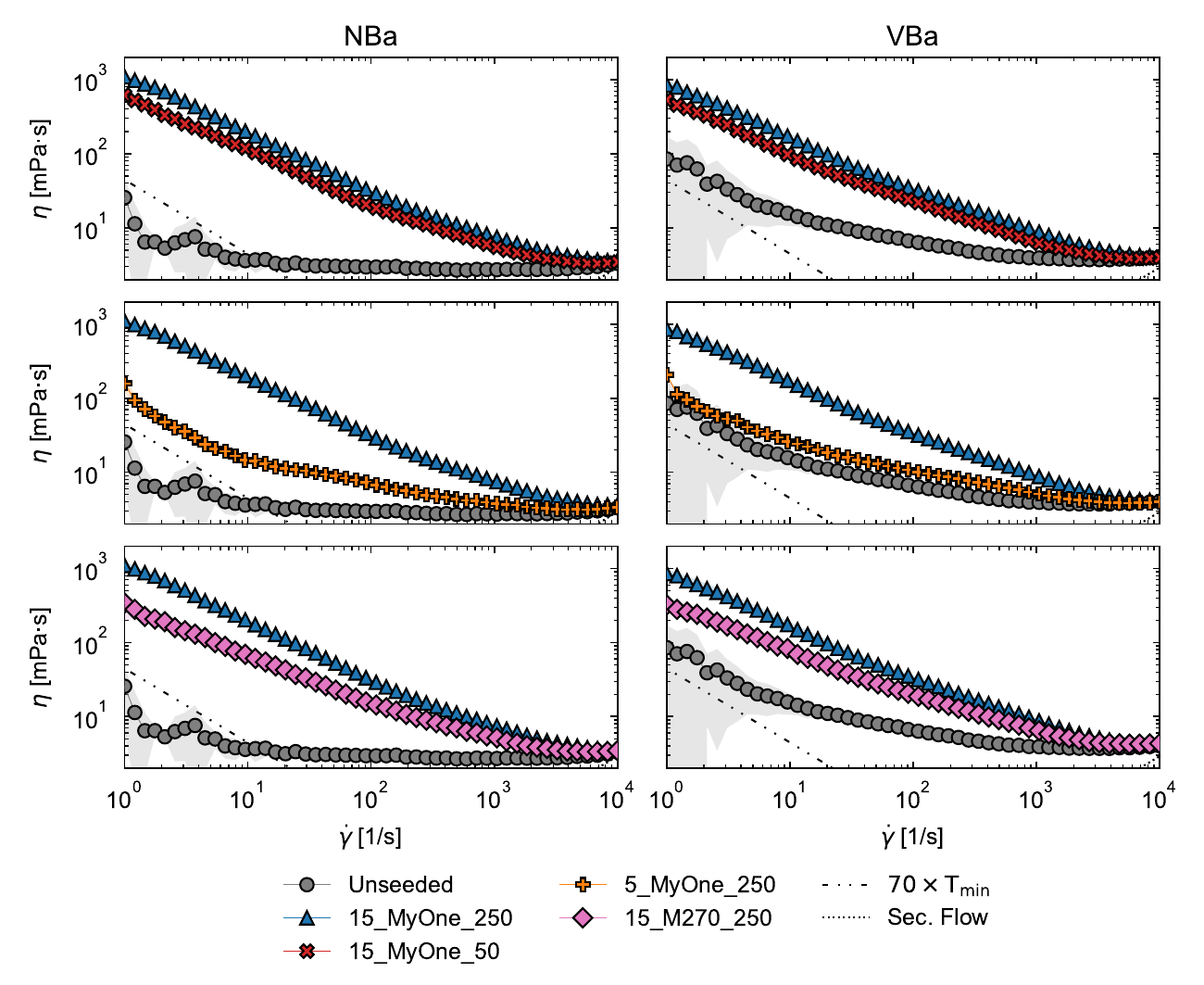}
\caption{Viscosity curves obtained with the NBa (left column) and VBa (right column), seeded with MyOne particles at 15 wt\% under the influence of 50 and 250 mT (top row), with MyOne particles at 5 and 15 wt\% under 250 mT (middle row), and with MyOne and M270 particles at 15 wt\% under 250 mT (bottom row). Adjusted low-torque and secondary flow limits are also shown in each graph.}
\label{fig:Serrated_flow_curves}
\end{figure}

Figure~\ref{fig:Serrated_flow_curves} shows the obtained viscosity curves for the different working fluids in Table~\ref{tab:working_fluids}. The columns of Figure~\ref{fig:Serrated_flow_curves} separate the samples with either analogue as the continuous phase, while the three rows compare the effects of altering the magnetic field density, the particle concentration and the particle type with the benchmark measurements and the unseeded analogues. Figure~\ref{fig:Serrated_flow_curves} also depicts the secondary flow\cite{turian1972} and adjusted low-torque\cite{ewoldt2015} limits, which are given by Equations~\ref{eq:sec_flow} and \ref{eq:low_torque}, respectively. Because the secondary flow limit is dependent on the gap height and the gap-error is significant, this limit was accordingly corrected ($h = h_\mathrm{c}+\varepsilon$).
\begin{equation}
    \text{Secondary flow}:\;\eta>\frac{\rho\dot\gamma h^3}{4R}\,,
    \label{eq:sec_flow}
\end{equation}
\begin{equation}
     \text{Adjusted minimum torque}:\;\eta>\frac{2\, (70\times\mathrm{T}_\mathrm{min})}{\pi R^3\dot\gamma}\,.
     \label{eq:low_torque}
\end{equation}

\begin{figure}[t]
\centering
\includegraphics[width=\linewidth]{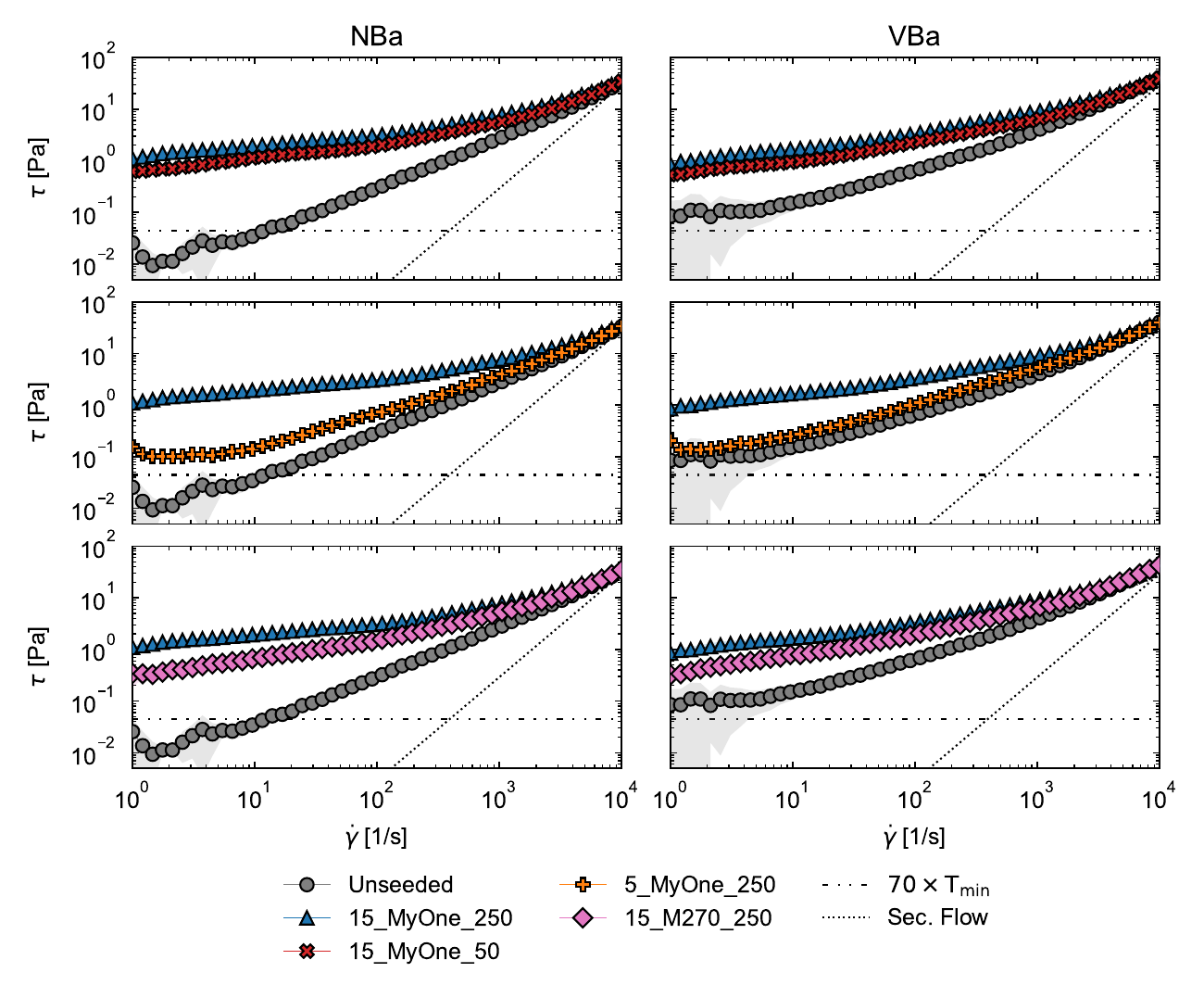}
\caption{Flow curves obtained with the NBa (left column) and VBa (right column), seeded with MyOne particles at 15 wt\% under the influence of 50 and 250 mT (top row), with MyOne particles at 5 and 15 wt\% under 250 mT (middle row), and with MyOne and M270 particles at 15 wt\% under 250 mT (bottom row). Adjusted low-torque and secondary flow limits are also shown in each graph.}
\label{fig:Stress_curves}
\end{figure}

First, considering the flow curves of the unseeded analogues, the NBa presents a reasonably constant viscosity of about 2.6 mPa$\cdot$s. In comparison, the VBa displays a clear shear-thinning behaviour with a high-shear viscosity of about 3.7 mPa$\cdot$s. Regarding the magnetorheological measurements, the carrier fluid does not introduce significant differences for a given set of particle concentration and magnetic field intensity (similar viscosity curves between the columns in Figure~\ref{fig:Serrated_flow_curves}). This could mean that the flow dynamics are dominated by the magnetic effects, which buffer the rheological differences between the continuous phases.

Comparing the unseeded analogues with the benchmark samples (15 wt\% of MyOne particles under 250 mT), there is a dramatic rheological alteration. The benchmark measurements display a striking shear-thinning behaviour with a significant low-shear viscosity increase ($\eta\approx1000$ mPa$\cdot$s at $\dot\gamma=1$ s\textsuperscript{-1}) which is due to the magnetic-induced structure of particle chains aligned with the field, perpendicular to the flow direction. Increasing the shear rate results in stronger viscous shearing forces that progressively break down the microstructure and consequently lead to a reduction of the measured viscosity. The small chains have practically no effect on the flow at high shear, and the measured viscosity approaches the non-magnetised suspensions'.

Reducing the magnetic field density from 250 to 50 mT did not result in a significant rheological alteration, but a slight overall viscosity reduction is noted. From the magnetization curves presented by \citeauthor{grob2018} we can gather that the particle magnetization is more significantly dependent on the field density for weaker fields, which could explain the subtle rheological alterations between 250 and 50 mT. On the other hand, reducing the particle concentration from 15 to 5 wt\% did lead to a considerable reduction in low-shear viscosity, with the 5 wt\% samples behaving almost similarly to the unseeded VBa. Regarding the particle type, the M270 is significantly larger (2.8 times larger than the MyOne), but the role of particle size is not yet clear\citep{deGans2000}. However, the M270 have a lesser magnetic susceptibility\citep{grob2018}, which is possibly the responsible parameter for the weaker magnetorheological response compared to the viscosity curve provided by the MyOne particles for the same concentration and magnetic field.

\begin{figure}[t]
\centering
\includegraphics[width=\linewidth]{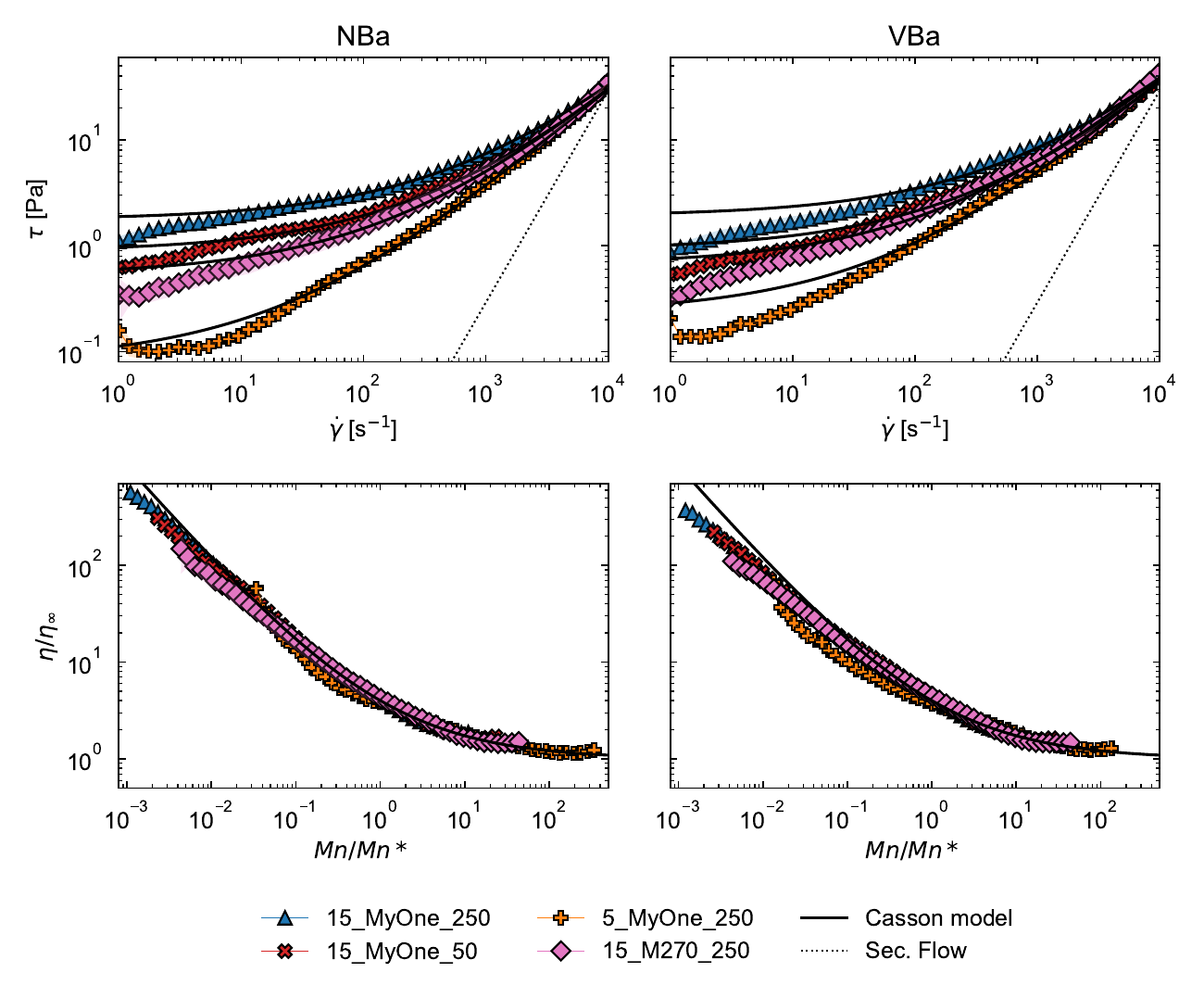}
\caption{Flow curves with Casson model fits (top row) and dimensionless viscosity as a function of the reduced Mason number (bottom row), for each tested particle type, concentration and magnetic field density combination, for samples with the NBa (left) and VBa (right) as continuous phase (unseeded data not shown). Secondary flow limit is shown along with the flow curves (top row).}
\label{fig:Master_curves}
\end{figure}

MR fluids are typically expected to have a yield stress. In Figure~\ref{fig:Stress_curves}, the flow curves (shear stress, $\tau$, against shear rate) are plotted, but for none of the tested samples does the stress clearly converge at low shear. With the 5wt\% MyOne samples and the unseeded VBa, there seems to be an apparent stress stabilization at low shear, but we believe this should be due to residual surface tension torque. This lack of yield stress evidence could mean that the particle concentration or the applied field strength are not sufficient to result in a yield behaviour, but a low-shear viscosity plateau is also not observable. We suspect that the apparent slip is responsible, even with the serrated geometry, and may be present on the smooth inferior plate. This is also corroborated by a small, characteristic bending of the flow curves. In any case, we can still estimate the yield stresses by fitting the Casson stress equation:

\begin{equation}
        \tau^{1/2} = \tau_\mathrm{y}^{1/2}+(\eta_\infty\dot\gamma)^{1/2}\,,
    \label{eq:Casson_stress}
\end{equation}
where $\tau_\mathrm{y}$ and $\eta_\infty$ are the yield stress and the large-shear viscosity, both of which are dependent on the particle magnetisation and the volume fraction. The fits were performed through a least-squares method on trimmed data sets ($10\leq\dot\gamma\leq7000$ s\textsuperscript{-1}), avoiding possible errors from apparent slip at low-shear and secondary flows at high-shear. Table \ref{tab:Casson_fit} gives the estimated yield stresses and low-shear viscosities, and the Casson model fitting curves are shown in the top row of Figure \ref{fig:Master_curves}. 

\begin{table}[htp]
    \centering
    \renewcommand{\arraystretch}{1.5}
    \small{
    \caption{Estimated yield stress and large-shear viscosity for each tested sample and magnetic field combination. Obtained through least-squares fit of the Casson model (Equation~\ref{eq:Casson_stress}) to the trimmed stress data ($10\leq\dot\gamma\leq7000$ s\textsuperscript{-1})}
    \label{tab:Casson_fit}
    \begin{tabularx}{\textwidth}{YY|YY|YY}
    \hline
    \multicolumn{2}{c|}{\centering Continuous phase}& 
    \multicolumn{2}{c|}{\centering NBa}&
    \multicolumn{2}{c}{\centering VBa}\\
    \hline
    \multicolumn{2}{c|}{\centering Fitting parameters}&
    $\tau_\mathrm{y}$ [Pa]&$\eta_\infty$ [mPa$\cdot$s]&
    $\tau_\mathrm{y}$ [Pa]&$\eta_\infty$ [mPa$\cdot$s]\\
    \hline
    \multicolumn{2}{c|}{\centering 15\_MyOne\_250}&1.764&1.939&1.908&2.304\\
    \multicolumn{2}{c|}{\centering 15\_MyOne\_50}&0.880&2.028&0.922&2.429\\
    \multicolumn{2}{c|}{\centering 5\_MyOne\_250}&0.080&2.724&0.230&3.081\\
    \multicolumn{2}{c|}{\centering 15\_M270\_250}&0.529&2.304&0.666&2.914\\
    \hline
    \end{tabularx}}
\end{table}

The Casson model seems to reasonably fit the experimental data at moderate to large shear rates, but diverges at low-shear, $\dot\gamma\lesssim10$ (even when fitting to the whole shear rate range\footnote{The Bingham model was also tested ($\tau=\tau_\mathrm{y}+\eta_\infty\dot\gamma$), but fit the data poorly for $\dot\gamma\lesssim100$ s\textsuperscript{-1}. Data shown in Supplementary Material, Section \ref{sec:sup:Bingham}.}), which was expected from the previously discussed lack of stress stabilisation. Nevertheless, the estimated yield stress values, despite quite low, seem to be representative of the magnetorheological strength, increasing with particle concentration and magnetic field density. Moreover, there is a general agreement between the tested continuous phases, with the VBa's returning slightly larger yield stress and large-shear viscosity values than the NBa's. The only significant discrepancy is found with the weakest magnetorheological samples, 5wt\% of MyOne particles and 250 mT, which may be more prone to slip due to the lack of a percolating structure.

An alternative analysis can be achieved through the adaptation of the Casson plastic equation (Equation \ref{eq:Casson_stress}) considering the Mason number, $Mn$, and its critical value, $Mn^*$ (which determines the transition from the magnetostatic to the hydrodynamic regime):
\begin{equation}
    Mn = \frac{72\eta_\mathrm{c}\dot\gamma}{\mu_0\mu_\mathrm{cr}(\rho_\mathrm{p}M)^2}\,,
    \label{eq:Mason_number}
\end{equation}
\begin{equation}
    Mn^* = \frac{72\eta_\mathrm{c}\tau_\mathrm{y}}{\mu_0\mu_\mathrm{cr}(\rho_\mathrm{p}M)^2\eta_\infty}\,.
    \label{eq:reduced_Mason_number}
\end{equation}
Here $\eta_\mathrm{c}$ is the continuous phase viscosity, $\mu_0$ is the vacuum permeability, $\mu_\mathrm{cr}$ is the continuous phase relative permeability ($\mu_\mathrm{cr}=1$), $\rho_\mathrm{p}$ is the particle density and $M$ is the particle magnetisation. A dimensionless form of the Casson equation can thus be written as\citep{morillas2019,berli2012}:
\begin{equation}
    \eta/\eta_\infty=1+(Mn/Mn^*)^{-1}+2(Mn/Mn^*)^{-1/2}\,.
    \label{eq:Master_curve}
\end{equation}
The bottom row of Figure \ref{fig:Master_curves} shows the obtained dimensionless viscosity curves along with the Casson model prediction given in Equation~\ref{eq:Master_curve}. The experimental data seems to reasonably collapse onto a single curve, as predicted by the Casson model. There are, however, significant deviations with the weakest magnetorheological tests (5 wt\% of MyOne particles and 250 mT). At high $Mn$ the viscosity curves follow the expected behaviour but drop-off at $Mn/Mn^*\lesssim1$. This effect is more clear with the VBa sample, whereas the NBa's viscosity seems to re-collapse onto the Casson model for $Mn/Mn^*\lesssim0.1$. In reality, this apparent viscosity increase ought to be related to low-shear errors, probably due to residual surface tension torque (see Figures \ref{fig:Serrated_flow_curves} and \ref{fig:Stress_curves}). With the remaining samples, a similar viscosity drop-off is also observable at very low Mason numbers ($Mn/Mn^*\lesssim0.03$), which could be due to apparent slip, as discussed before.

\subsection{Oscillatory shear}

\subsubsection{Unseeded analogues}

\begin{figure}[htp]
    \centering
    \begin{minipage}{\textwidth}
    \centering
    Unseeded
    \end{minipage}
    \centering
    \begin{minipage}{\textwidth}
    \centering
    \small
    \includegraphics[trim={0cm 1cm 0cm 1.5cm},clip,width=\textwidth]{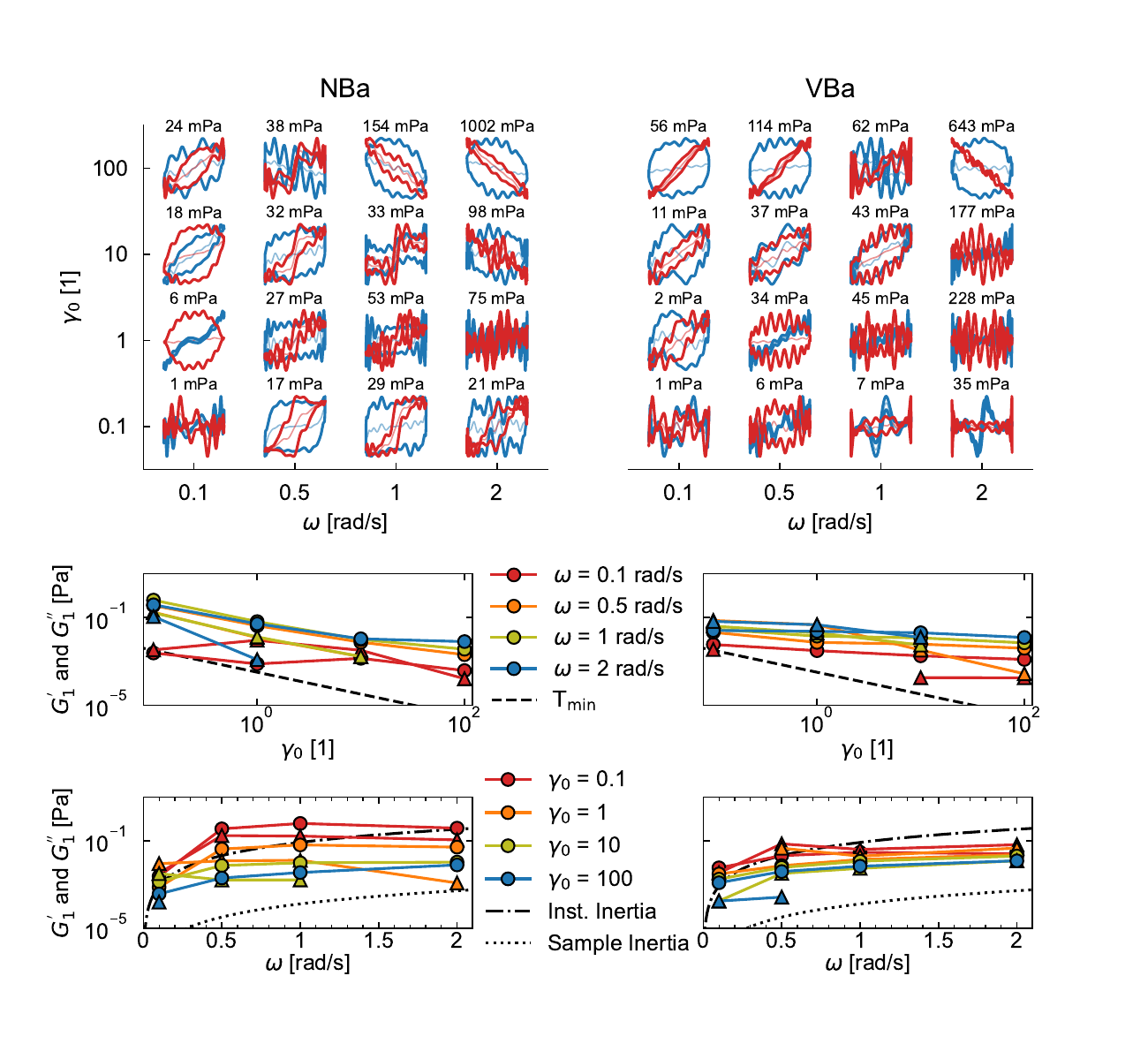}
    \end{minipage}
    \caption{Oscillatory shear data gathered with the unseeded NBa (left) and VBa (right), with the PP20 MRD P2. (Top) Pipkin diagrams depicting the Lissajous curves. The blue and red curves represent, respectively, the elastic (stress vs strain) and viscous (stress vs strain rate) Lissajous curves. The elastic and viscous stresses are also lightly plotted in the respective colours and the maximum measured stress is presented above each plot. (Bottom) First-harmonic loss ($G''_1$, in circular markers) and storage ($G'_1$, in triangular markers) moduli and experimental limits associated with low-torque issues and instrument and sample inertia (corrected for the geometry's gap-error).}
    \label{fig:Oscillatory_PP20_MRD_P2_0wt_0mT}
\end{figure}

Figure~\ref{fig:Oscillatory_PP20_MRD_P2_0wt_0mT} shows the Pipkin diagrams\citep{pipkin2012} of Lissajous curves and the first-harmonic viscoelastic moduli\footnote{Moduli data was deleted if specific error messages returned from the rheometer software regarding either the viscous or elastic torque, respectively.} for the unseeded samples. The elastic (stress vs strain, in blue) and viscous (stress vs strain rate, in red) Lissajous curves show the raw stress response and are max-normalised (adimensional). The maximum measured stress is presented above the plots and the decoupled elastic and viscous stresses\citep{cho2005} are also lightly plotted.

Focusing on the Lissajous curves (top of Figure~\ref{fig:Oscillatory_PP20_MRD_P2_0wt_0mT}), the data is clearly ridden with experimental error. Since the NBa is Newtonian, the expected response would be a perfectly circular elastic curve and a positive-diagonal of the viscous response. At low frequency and moderate strain ($\omega = 0.1$ rad/s, $1\leq\gamma_0\leq10$) the NBa seems to present a viscoelastic response that could be due to additional torque from surface tension forces while at large frequency and amplitude ($1\leq\omega\leq2$ rad/s, $\gamma_0=100$) the stress is in opposite phase with the strain rate, which could be a symptom of severe inertial effects. With the VBa, at low frequency and large strain ($0.1\leq\omega\leq0.5$ rad/s, $\gamma_0=100$) there is a semblance of the expected subtle viscoelasticity. Still, the remaining data is so deeply affected by experimental error that it is not possible to ascertain whether this represents the fluid's actual rheology. 

As we have discussed, the first-harmonic moduli provide limited information, but they are still useful to plot against known experimental limits, allowing to evaluate the measurement window. The low-torque limit, for a Parallel-Plate geometry, is given by\citep{ewoldt2015}:
\begin{equation}
    G>\frac{2 \mathrm{T}_\mathrm{min}}{\pi R^3 \gamma_0}\,,
\end{equation}
where $G$ is either the loss ($G''$) or storage modulus ($G'$), $\mathrm{T}_\mathrm{min}$ is the instrument's minimum torque in oscillatory shear and $R$ is the geometry radius. At high frequencies, inertial limitations become important. On the one hand, the instrument inertia limit is given by\citep{ewoldt2015}:
\begin{equation}
    G>I\frac{2h}{\pi R^4}\omega^2\,,
\end{equation}
where $I$ is the geometry's moment of inertia and $h$ is the gap between plates. On the other hand, the sample inertia limit can be written as\citep{ewoldt2015}:
\begin{equation}
    |G^*|>\left(\frac{10}{2\pi}\right)^2\rho\omega^2h^2\,,
\end{equation}
where $|G^*| = \sqrt{G'^2+G''^2}$ is the magnitude of the complex modulus. Similar to the secondary flow limit for steady shear, we correct the inertia limits for the gap-error of the PP20 MRD P2. These experimental limits are also shown in Figure~\ref{fig:Oscillatory_PP20_MRD_P2_0wt_0mT} (bottom). As can be seen, the reliable experimental window is very small. On the one hand, we approach the low-torque limit at small amplitudes, which is not corrected by any safety coefficient to account for surface tension torque (contrarily to the steady-shear case). At high frequencies, inertial issues, also enlarged by the gap-error, become problematic with the instrument inertia limit always practically overrun. Therefore, little information can be gathered regarding which limitation is responsible for which result.

What we can conclude, however, is that the PP20 MRD P2 is unsuitable for characterising the unseeded analogues' rheological properties under oscillatory shear, at least within the tested experimental range. Looking at the expressions for the experimental limitations, the most helpful way to widen the experimental window is to increase the geometry diameter. We performed the same oscillatory shear measurements with a larger, smooth PP50 (Parallel-Plate with 50 mm of diameter) in an additional bottom plate (the magnetorheological cell has a geometrical constraint of approximately 30 mm on standard geometries and 20 mm on the MRD geometries) and found a significant improvement of the result quality. At large strain ($\gamma_0=100$) we encountered the expected purely-viscous NBa response, and evidence of non-linear viscoelasticity with the VBa. However, because our magnetorheological setup does not permit such measuring geometries, we present these results as supplementary material (Section \ref{sec:sup:PP50}).

\subsubsection{Magnetorheological benchmark measurements}

\begin{figure}[htp]
    \centering
    \begin{minipage}{\textwidth}
    \centering
    15\_MyOne\_250
    \end{minipage}
    \centering
    \begin{minipage}{\textwidth}
    \centering
    \small
    \includegraphics[trim={0cm 1cm 0cm 1.5cm},clip,width=\textwidth]{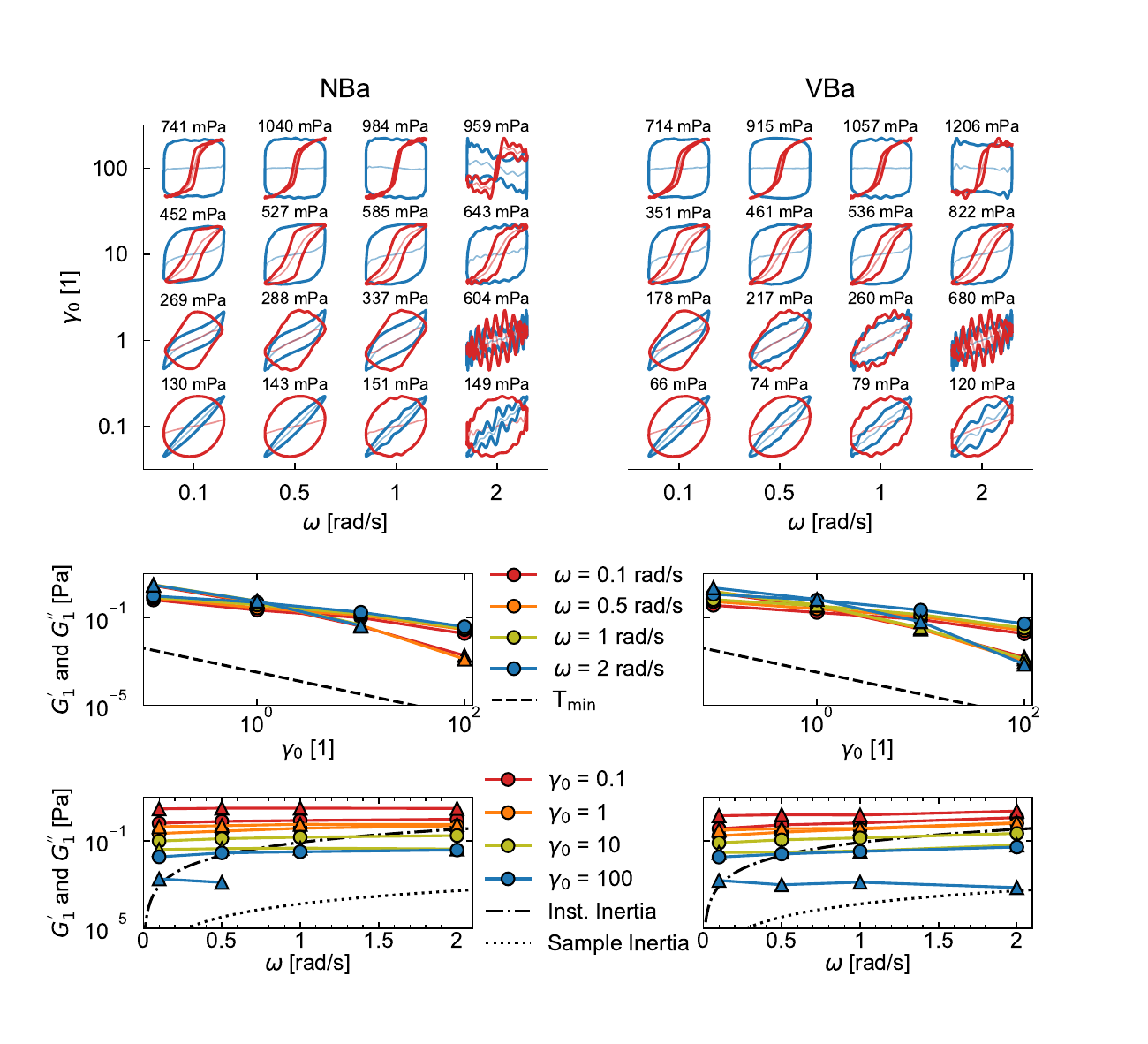}
    \end{minipage}
    \caption{Oscillatory shear data gathered with the NBa (left) and VBa (right) seeded with 15 wt\% of MyOne particles under a 250 mT magnetic field, with the PP20 MRD P2. (Top) Pipkin diagrams depicting the Lissajous curves. The blue and red curves represent, respectively, the elastic (stress vs strain) and viscous (stress vs strain rate) Lissajous curves. The elastic and viscous stresses are also lightly plotted in the respective colours and the maximum measured stress is presented above each plot. (Bottom) First-harmonic loss ($G''_1$, in circular markers) and storage ($G'_1$, in triangular markers) moduli and experimental limits associated with low-torque issues and instrument and sample inertia (corrected for the geometry's gap-error).}
    \label{fig:Oscillatory_15wt_MyOne_250mT}
\end{figure}

Figure~\ref{fig:Oscillatory_15wt_MyOne_250mT} shows the obtained Pipkin diagrams and viscoelastic moduli plots of the benchmark measurements (15 wt\% of MyOne particles under the effects of a 250 mT magnetic field density). Across the whole frequency/amplitude range the rheological differences between the continuous phases do not appear, as the results of both the seeded NBa and VBa samples are very similar, aligning with the steady shear findings. For the largest tested frequencies, $\omega=1$ and 2 rad/s, there are oscillations in the Lissajous curves that may be due to inertial issues. Compared to the data gathered with the unseeded analogues, there is a significant increase of the viscoelastic moduli across the tested range which allows to escape from the experimental limits, meaning the setup can capture the magnetorheological properties of the seeded samples, at least for this particle type, concentration and applied magnetic field density. 

With frequency increase, the elastic response is slightly reduced. Still, within the tested range, the magnetorheological behaviour seems to be practically frequency-independent, which is expected at low frequency\citep{parthasarathy1999}. At low amplitude, $\gamma_0 = 0.1$, the response is elastic-dominated, with elliptical Lissajous and a slight bending of the viscous stress which points toward a degree of non-linearity. With strain increase the response becomes progressively more viscous (moduli crossover between $1\lesssim\gamma_0\lesssim10$) and non-linear. At high strain, $\gamma_0=100$, the elastic contribution is practically negligible and the Lissajous curves display a viscoplastic response.

\begin{figure}[htp]
    \centering
    \begin{minipage}{\textwidth}
    \centering
    15\_MyOne\_250 ($\omega = 0.1$ rad/s)
    \end{minipage}
    \centering
    \begin{minipage}{\textwidth}
    \centering
    \small
    \includegraphics[width=\textwidth]{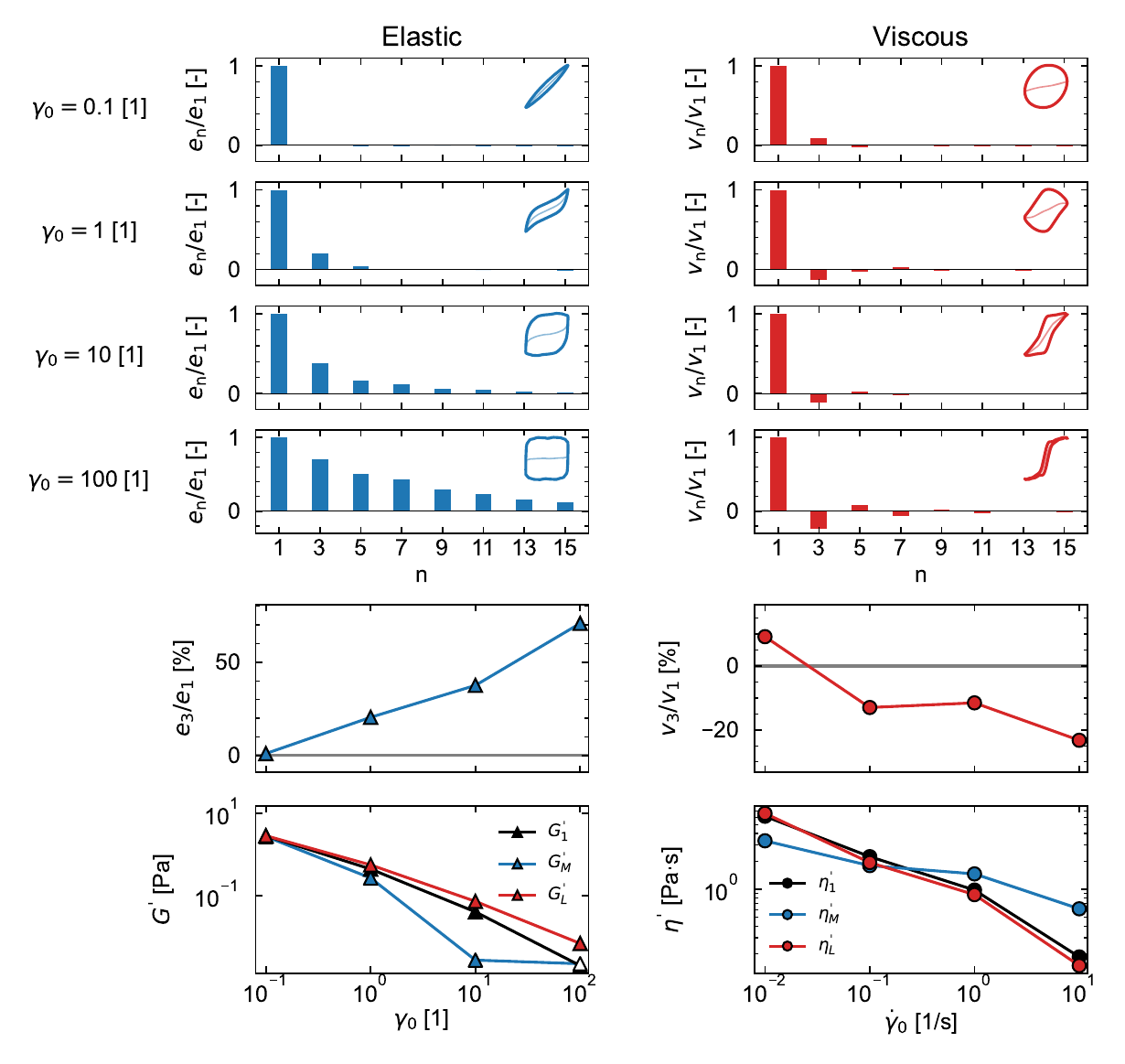}
    \end{minipage}
    \caption{Oscillatory shear data gathered with the NBa with 15 wt\% of MyOne particles under a 250 mT magnetic field, with the PP20 MRD P2 ($\omega=0.1$ rad/s). Scaled Chebyshev coefficient spectrum, $e_\mathrm{n}/e_1$ and $v_\mathrm{n}/v_1$ (with reconstructed Lissajous curves), variation of the scaled third harmonic coefficients, $e_3/e_1$ and $v_3/v_1$, and relevant viscoelasticity measures (elastic: $G_1'$, $G_M'$, $G_L'$, and viscous: $\eta_1'$, $\eta_M'$, $\eta_L'$) with applied deformation/deformation rate (negative $G_M'$ values are presented as open markers). Analysed with MITlaos\citep{mitlaos}.}
    \label{fig:15wt_MyOne_250mT_MITlaos}
\end{figure}

The moduli plots of Figure~\ref{fig:Oscillatory_15wt_MyOne_250mT} show, from the decrease of either $G'_1$ and $G''_1$, an average elastic softening and viscous thinning, but the intricacies of the response's non-linearity remain obscure (the Lissajous curves do shed some light on the non-linear response, albeit qualitatively). We used the MITlaos software\citep{mitlaos} to quantitatively describe the sample's non-linear magnetorheological response through FT-Chebyshev analysis. Because there seem to be no behavioural differences between the samples with either analogue, nor frequency-dependence apart from seemingly-inertial errors at large frequency, we will focus on the results obtained with the NBa benchmark sample (15 wt\% of MyOne under 250 mT) at the minimum driving frequency ($\omega=0.1$ rad/s). The raw stress and strain data were imported to MITlaos, and the signal was filtered for the first 15 harmonics. Figure~\ref{fig:15wt_MyOne_250mT_MITlaos} exhibits the obtained viscoelastic moduli relevant in the non-linear regime ($G_1'$, $G_M'$, $G_L'$ and $\eta_1'$, $\eta_M'$, $\eta_L'$), the Chebyshev coefficient spectrum (until the 15\textsuperscript{th} harmonic) with the corresponding reconstructed Lissajous curves, and the relative contribution of the third-order coefficients.

Considering the elastic measures (left column of Figure~\ref{fig:15wt_MyOne_250mT_MITlaos}), at the minimum strain, $\gamma_0 = 0.1$, all elastic moduli approach the same value, $G_M' \approx G_L' \approx G_1'$, relating to a linear elastic response also showcased by the lack of high-order contributions on the elastic Chebyshev spectrum. Increasing the strain leads to a divergence of $G_M'$ and $G_L'$, with the first decreasing faster. This onset of elastic non-linearity can also be observed by the appearance of high-order Chebyshev coefficients, which become progressively more significant with input-strain increase. Notably, at the maximum imposed deformation ($\gamma_0=100$), the minimum strain modulus takes on a negative value, $G_M'\approx-2.22\times10^{-3}$ Pa (negative values of $G_M'$ are presented as open markers), corresponding to a negative slope of the elastic Lissajous at $\gamma=0$. Negative values of $G_M'$ are associated with secondary loops of the viscous Lissajous curve in a strong non-linear response where the sample is discharging elastic stress at a faster rate than the deformation is being applied, and are expected at large strain and frequency\citep{ewoldt2010}. Looking at the Lissajous curves in question, there are no prominent secondary loops on the viscous response and the negative slope of the elastic stress appears to be a result of local oscillations of the stress at $\dot\gamma\to\dot\gamma_0$, which should be due to either inertial or low-torque issues as the sample softens. The third-order Chebyshev coefficient contribution is always positive in the NLVE ($e_3>0$), meaning the strain rate drives the average elastic softening.

Regarding the viscous properties (right column of Figure~\ref{fig:15wt_MyOne_250mT_MITlaos}), already at the minimum tested rate ($\dot\gamma_0=0.01$ s\textsuperscript{-1}) $\eta_L'>\eta_M'$ and a non-negligible positive contribution of $v_3$ is observable, meaning the viscous non-linearity onsets before the elastic. At $\dot\gamma_0=0.1$ s\textsuperscript{-1}, $\eta_M'\approx\eta_L'$,  which could point towards a return to linear behaviour had we not also evaluated the Lissajous and the Chebyshev coefficients, which clearly display a non-linear response. There is a significant transition between $0.01<\dot\gamma_0<0.1$ s\textsuperscript{-1} (or $0.1<\gamma_0<1$), with $v_3$ changing from positive to negative, which points towards significant microstructural changes.

\begin{figure}[htp]
    \centering
    \begin{minipage}{\textwidth}
    \centering
    15\_MyOne\_250 ($\omega = 0.1$ rad/s)
    \end{minipage}
    \centering
    \begin{minipage}{\textwidth}
    \centering
    \small
    \includegraphics[trim={0cm .5cm 0cm 1cm},clip,width=\textwidth]{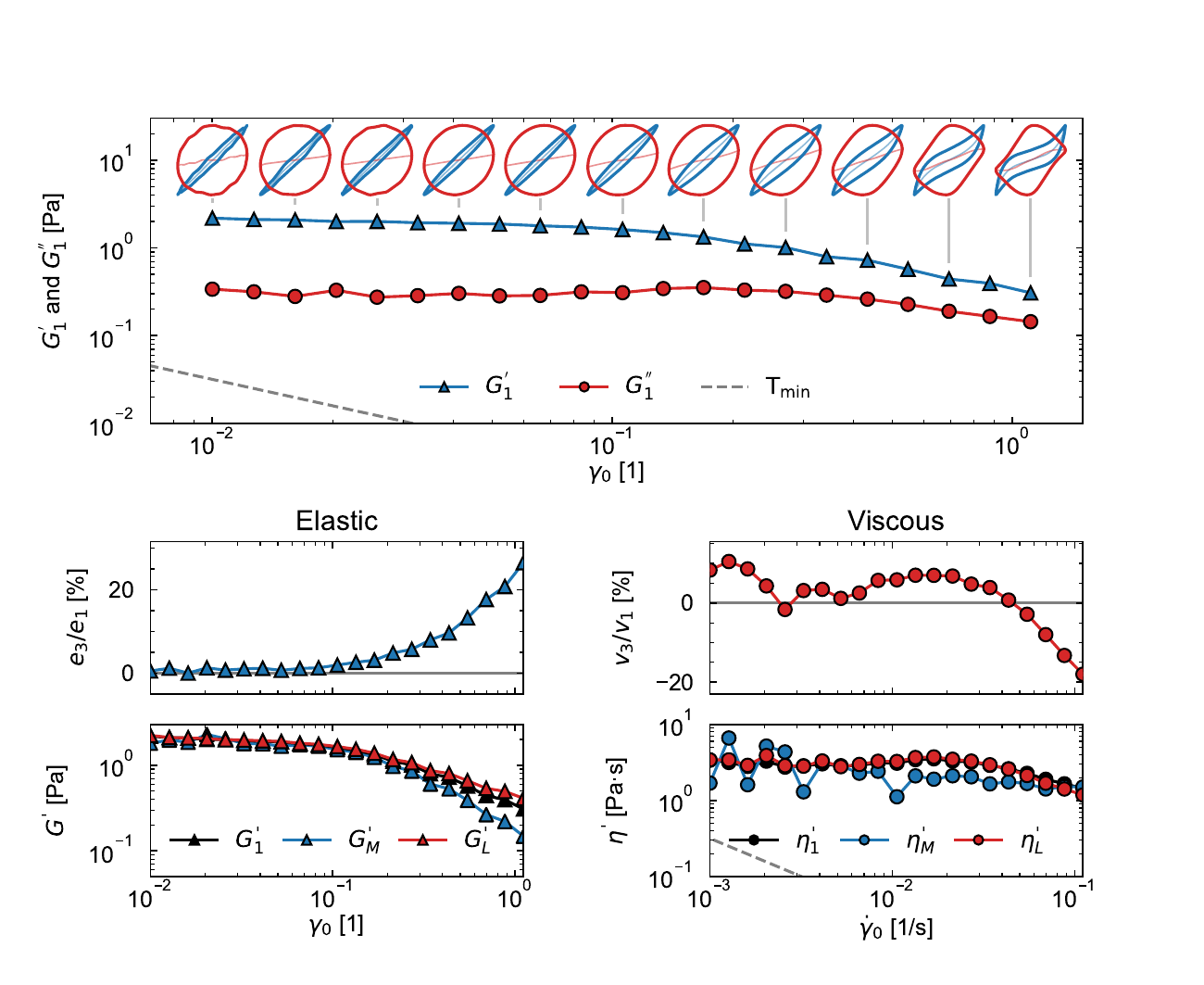}
    \end{minipage}
    \caption{Oscillatory shear data gathered with the NBa with 15 wt\% of MyOne particles under a 250 mT magnetic field, with the PP20 MRD P2 ($\omega=0.1$ rad/s). (Top) first-harmonic loss ($G''_1$, in circular markers) and storage ($G'_1$, in triangular markers) moduli and elastic and viscous Lissajous curves (reconstructed for the first 15 harmonics), in blue and red respectively, and corresponding stresses (lightly plotted). (Bottom) Scaled 3\textsuperscript{rd}-order Chebyshev coefficients, $e_3/e_1$ and $v_3/v_1$, and relevant viscoelasticity measures (elastic: $G_1'$, $G_M'$, $G_L'$, and viscous: $\eta_1'$, $\eta_M'$, $\eta_L'$). Data from MITlaos\citep{mitlaos}.}
    \label{fig:Benchmark_depth}
\end{figure}

To further probe this microstructural transition, we performed a finer amplitude sweep at $\omega=0.1$ rad/s, also testing lesser strains to better locate the LVE. Figure~\ref{fig:Benchmark_depth} shows the first-harmonic loss and storage moduli and some corresponding Lissajous curves, as well as the relevant quantities in the non-linear regime: elastic moduli, dynamic viscosities and third-order Chebyshev coefficients from MITlaos. The LVE seems to extend up to $\gamma_0\approx0.03$ (where $G'_1\approx0.9\,G'_1(\gamma_0\to0)$), with an elastic-dominant response (solid-like). At very small strains, oscillations of the Lissajous curves and viscous data scattering, particularly of $v_3/v_1$ and $\eta'_M$, point towards low-torque issues. The elastic measures, less affected by the error, show a constant, null contribution of $e_3$ and $G'_M\approx G'_L\approx G'_1$.

Entering the NLVE, and considering the elastic response first, before any significant contribution from high-order harmonics, a slight decrease of $G'_1$ is associated to first-harmonic non-linearities\citep{ewoldt2013}. From $\gamma_0\gtrsim0.1$, the decrease of $G'_1$ is accelerated, accompanied by an increasing positive contribution of $e_3$ and a faster decrease of $G'_M$. Throughout the entire tested strain range, $G_1'$ decreases, with $e_3>0$ and $G'_M>G'_L$, which relates to an elastic softening of the sample driven by the strain rate (lesser elastic stress than the linear basis at $\dot\gamma\to\dot\gamma_0$).

Regarding the viscous non-linearities, In the beginning of the NLVE we observe the viscous non-linearity onseting first, at $\dot\gamma_0\approx0.005$ s\textsuperscript{-1} ($\gamma_0\approx0.05$), with an increase of $v_3/v_1$ and $\eta'_L$ that should be associated with the overshoot of $G''_1$ (type III behaviour\citep{hyun2002}). As such, the viscous non-linearity begins with viscous thickening driven by the strain rate, returning larger viscous stresses than the linear baseline at $\dot\gamma\to\dot\gamma_0$, where the microstructural rearrangements occur. Increasing the strain rate we observe the simultaneous decrease of $G''_1$, $\eta'_L$ and $v_3/v_1$, with the latter switching signs at $\dot\gamma_0\approx0.05$ s\textsuperscript{-1} ($\gamma_0\approx0.5$). This is indicative of a transition from viscous thickening to thinning, both driven by the strain rate, which ought to be associated with the onset of large intracycle structural modifications and, at some point, the beginning of microstructural breakdown at $\dot\gamma\to\dot\gamma_0$.

The peak of $G''_1$ and $v_3/v_1$ are simultaneous, meaning the onset of viscous thinning (decrease of $G''_1$) occurs when the contribution of $v_3$ is still positive. This could be due to the non-constant shear rate applied in PP geometries that leads to different microstructural responses along the geometry radius, thus having no significant microstructural relevance, or it could signify a momentary strain-dependence of the non-linearity. Considering the second hypothesis, due to the time-dependent nature of oscillatory shear, it could be possible that for a small strain, chain rotation is sufficient to comply with the applied deformation, independently of the strain rate, given that the oscillatory frequency is small. As such, only beyond a critical strain would the microstructure begin to breakdown due to viscous shearing forces. We performed a short numerical work to understand the mechanism behind the beginning of viscous thinning.

\subsection{Numerical analysis}

\begin{figure}[htp]
    \centering
    \small
    \includegraphics[width=.9\textwidth]{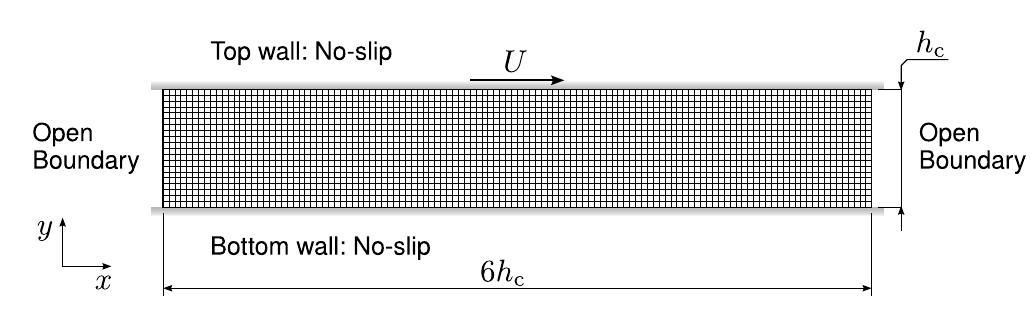}
    \caption{Schematic of the numerical model displaying the employed mesh, boundary conditions and fluid velocity imposition at the top wall ($U$).}
    \label{fig:Numerical_model_schematic}
\end{figure}

We set out to model the response of a magnetised-particle chain under oscillatory shear in COMSOL MultiPhysics, seeking to gather information on the critical strain required to provoke chain subdivision. The three-dimensional rotational flow imposed by the rheometer was approximated by a two-dimensional flow in a straight rectangular channel ($h_\mathrm{c}\times6h_\mathrm{c}$, with $h_\mathrm{c} =0.1$ mm) driven by a velocity of the top channel wall, $U$. The domain was meshed with 2400 equal-sized quadrilateral elements (of side $l=h_\mathrm{c}/20$)\footnote{For the sake of conciseness, we present a mesh analysis as supplementary material, Section \ref{sec:sup:mesh}}. The flow was calculated by solving the equations of conservation of mass and momentum for the isothermal, incompressible and laminar flow of a Newtonian fluid:
\begin{equation}
\rho\left(\frac{\partial\mathbf{u}}{\partial t}+\mathbf{u}\cdot\mathbf{\nabla}\right)\mathbf{u}=-\mathbf{\nabla}p+\eta\mathbf{\nabla}^2\mathbf{u}+\mathbf{F}\,,
\end{equation}
\begin{equation}
    \mathbf{\nabla}\cdot\mathbf{u}=0\,.
\end{equation}
No-slip boundary conditions were imposed at the top and bottom walls, and open-boundary conditions were applied at the lateral walls, allowing free flow in and out of the channel. Figure \ref{fig:Numerical_model_schematic} presents a schematic of the numerical model.

\begin{figure}[htp]
    \centering
    \small
    \includegraphics[width=0.4\textwidth]{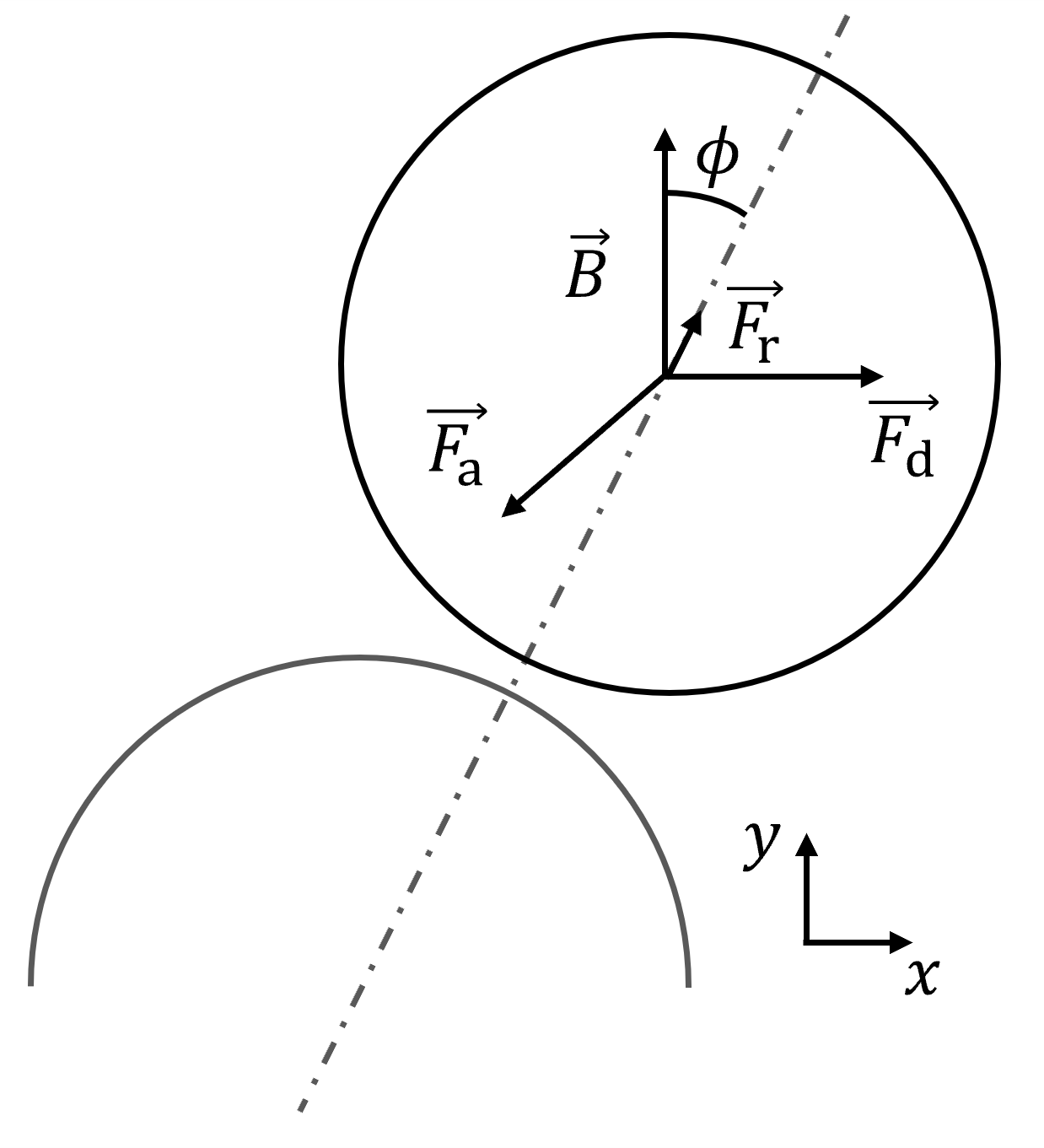}
    \caption{Schematic of forces acting on a particle of a two-element chain (particles slightly distanced, not to scale). Flow is left to right (positive $x$) and magnetic field, $\vec{B}$, oriented vertically (positive $y$).}
    \label{fig:Particle_forces_schematic}
\end{figure}

The input fluid properties were those of the NBa ($\rho = 1050$ kg/m\textsuperscript{3} and $\eta = 2.6$ mPa$\cdot$s). The particles ($\rho_\mathrm{p} = 1800$ kg/m\textsuperscript{3}, equal the MyOne) were released at the centre of the channel, evenly distributed along the gap, and the magnetic-dipole forces were modelled as interparticle interactions according to the formulation of \citeauthor{melle2003} Particle overlapping was avoided through an excluded-volume force, such that particles aligned with the field and distanced $1.1\,d_\mathrm{p}$, suffered a repulsive force about 14 times smaller than the magnetic attraction, reducing to zero when the particles are in contact\citep{gao2012}. This force also facilitates convergence, slowing the collision between approaching particles. To avoid needlessly large simulations, the particle size was increased and two values were tested, $d_\mathrm{p}=10$ and $25$ \textmu m, resulting in two gap-spanning chain sizes ($l_\mathrm{chain} = \lfloor h_\mathrm{c}/d_\mathrm{p}\rfloor$=10 and 4, respectively)\footnote{Particle/solid contact is computed at the particle centre.}. Moreover, the particle magnetisation was reduced to allow more reasonable time steps: $M^* \approx 1.88\times10^{-1}$ Am\textsuperscript{2}/kg (50 times lesser than the MyOne saturation magnetisation\citep{grob2018}), and, lastly, here we focus on the influence of the fluid flow on the chain dynamics and the inverse phenomenon is disregarded, which allows us to solve for the flow field and particle dynamics separately, but impedes gathering information regarding the influence of the chain dynamics on the viscous stress felt at the solid boundaries. Particle sedimentation, Brownian motion and inertial effects are disregarded on the same basis as in our previous work\citep{rodrigues2025} and, therefore, the forces acting on the particles are the viscous drag ($  \boldsymbol{F_\mathrm{d}}$) and the interparticle magnetic attraction ($\boldsymbol{F_\mathrm{a,ij}}$) and repulsive excluded-volume force ($ \boldsymbol{F_\mathrm{r,ij}}$), given by, respectively:
\begin{equation}
    \boldsymbol{F_\mathrm{d}} = 3\,\pi\,\eta\,d_\mathrm{p}(\boldsymbol{u_\mathrm{c}}-\boldsymbol{u_\mathrm{p}})\,,
\end{equation}
\begin{equation}
    \boldsymbol{F_\mathrm{a,ij}} = \frac{3\,\mu_0\,\mu_\mathrm{cr}\,m^2}{4\,\pi\,r_\mathrm{ij}^4}\left[\left(1-5\left(\hat{m}\cdot\hat{r}_\mathrm{ij} \right)^2\right)\hat{r}_\mathrm{ij}+2\left(\hat{m}\cdot\hat{r}_\mathrm{ij} \right)\hat{m}   \right]\,,
\end{equation}
\begin{equation}
    \boldsymbol{F_\mathrm{r,ij}} = 2\,\frac{3\,\mu_0\,\mu_\mathrm{cr}\,m^2}{4\,\pi\,d_\mathrm{p}^4}\exp\left[{-30\left(\frac{r_\mathrm{ij}}{d_\mathrm{p}}-1\right)}\right]\hat{r}_\mathrm{ij}\,,
\end{equation}
where $(\boldsymbol{u_\mathrm{c}}-\boldsymbol{u_\mathrm{p}})$ is the particle velocity ($\boldsymbol{u_\mathrm{p}}$) relative to the continuous phase ($\boldsymbol{u_\mathrm{c}}$), $\mu_0$ is the vacuum permeability, $\mu_\mathrm{cr}$ is continuous phase relative permeability ($\mu_\mathrm{cr}=1$), $\boldsymbol{r_\mathrm{ij}}$ is the interparticle vector (magnitude $r_\mathrm{ij}$ and direction $\hat{r}_\mathrm{ij}$) and $\boldsymbol{m_\mathrm{ij}}$ is the magnetic moment (with magnitude $m=4/3\pi(d_\mathrm{p}/2)^3\,\rho_\mathrm{p}\,M^*$ and direction $\hat{m}_\mathrm{ij}$, perpendicular to the flow direction). In Figure~\ref{fig:Particle_forces_schematic}, a schematic of the forces applied to a representative 2-particle chain is shown. Regarding the temporal discretization, we employed an adaptive scheme, allowing COMSOL to adjust the time step freely, except for the initial value of the particle simulation, which was defined very small to guaranty convergence ($\Delta t_\mathrm{0} = 10^{-6}$\,s).

\begin{figure}[htp]
    \centering
    \small
    \includegraphics[width=\textwidth]{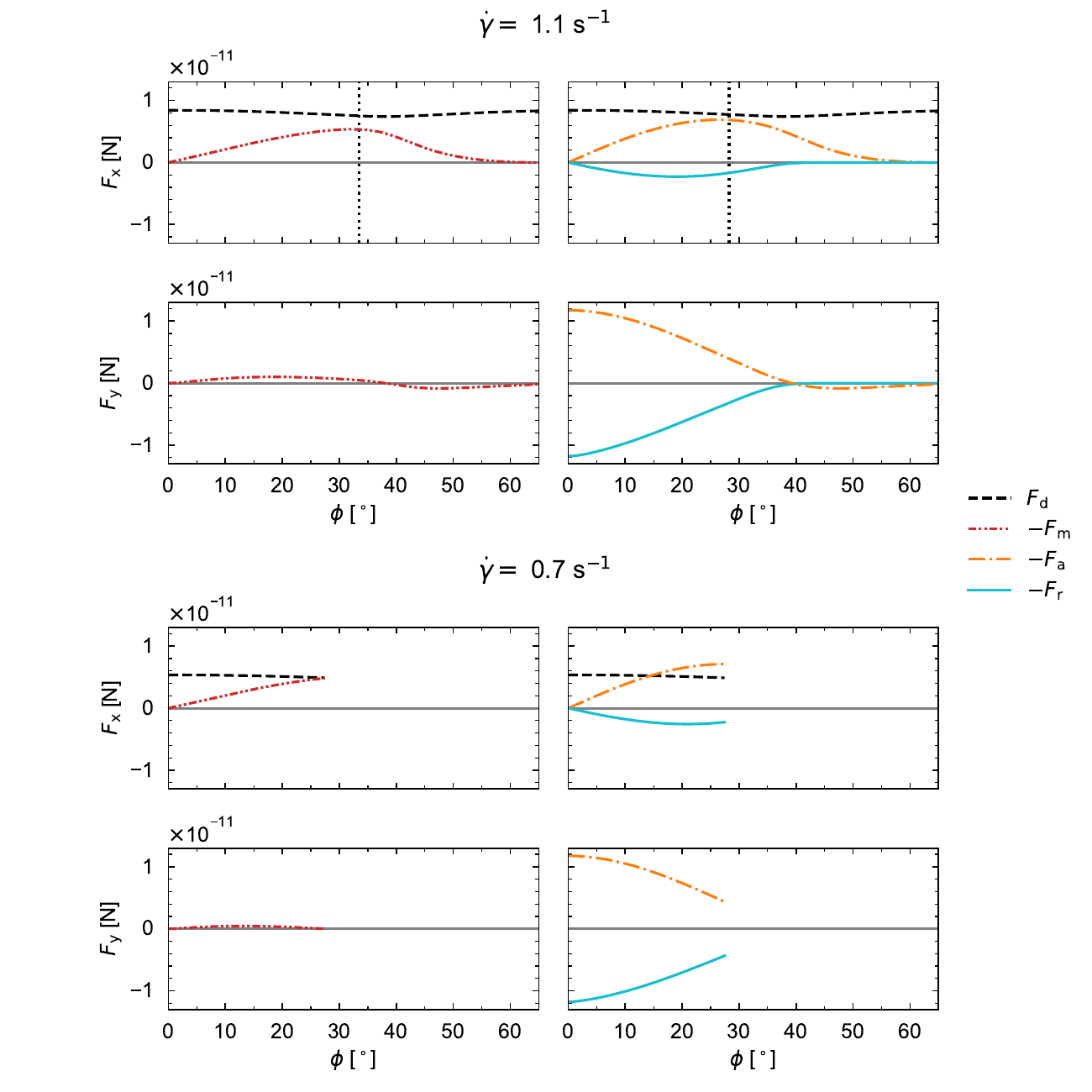}
    \caption{Forces acting on a 2-element chain under steady shear against the chain angle, $\phi$. Shear rate (top) $\dot\gamma = 1.1$ and (bottom) $0.7$ s\textsuperscript{-1}. Acting forces are: viscous drag, $\boldsymbol{F_\mathrm{d}}$, magnetic attraction, $\boldsymbol{F_\mathrm{a}}$, repulsive excluded-volume force, $\boldsymbol{F_\mathrm{r}}$, and global magnetic force, $\boldsymbol{F_\mathrm{m}} = \boldsymbol{F_\mathrm{a}}+\boldsymbol{F_\mathrm{r}}$, as displayed in Figure~\ref{fig:Particle_forces_schematic}.}
    \label{fig:Critical_chain_angle}
\end{figure}

Before proceeding with the oscillatory shear simulations, we tested for steady shear first to evaluate the critical shear rates for chain subdivision and complete breakdown. The shear rate was varied by imposing a top wall velocity $U = \dot\gamma\, h_\mathrm{c}$. It was found that both chains breakdown to individual particles when $\dot\gamma_\mathrm{brk.}\approx0.79$ s\textsuperscript{-1}, which was expected because chain breakdown is equivalent to the subdivision of any two-element chain. On the other hand, for $l_\mathrm{chain} = 4$ the critical subdivision shear rate was $\dot\gamma_\mathrm{sub.}\approx 0.23$ s\textsuperscript{-1}, while the largest chain with $l_\mathrm{chain} = 10$ subdivided earlier, at $\dot\gamma_\mathrm{sub.}\approx0.04$ s\textsuperscript{-1} (data not shown here for conciseness). Still considering steady shear, we can also gather information on the critical angle for chain break, $\phi_\mathrm{c}$. It is relevant, however, that while in reality the particle contact force is a step function triggered for $r_\mathrm{ij}=d_\mathrm{p}$, the excluded-volume force in our formulation is significant for small interparticle distances and a global magnetic force can be computed as $\boldsymbol{F_\mathrm{m,ij}} = \boldsymbol{F_\mathrm{a,ij}} +\boldsymbol{F_\mathrm{r,ij}}$. In Figure~\ref{fig:Critical_chain_angle} are plotted the forces acting on a 2-element chain subject to a steady flow with shear rate $\dot\gamma=1.1$ and 0.7 s\textsuperscript{-1} against the chain angle, $\phi$. On the left column of Figure~\ref{fig:Critical_chain_angle} are shown the global magnetic and viscous drag forces, and on the right column, the magnetic attraction and repulsive excluded-volume forces are shown decoupled (see Figure~\ref{fig:Particle_forces_schematic} for reference). 

We can see that the viscous shearing force weakens slightly as the chain rotates due to the reduction of the vertical disparity between the particles ($\Delta y = r_\mathrm{ij}\,\cos(\phi)$). Contrarily, the magnetic force increases on both axes ($x$ and $y$) as the magnetic dipole interactions try to re-align the chain with the magnetic field (aligned with $y$). If this simultaneous decrease of the relative drag and increase of the magnetic force is sufficient to reach an equilibrium ($F_\mathrm{d,x}+F_\mathrm{m,x}=0$) the chain remains stable, as is the case for $\dot\gamma=0.7$ s\textsuperscript{-1}. However, if, despite the force convergence, the viscous drag remains too strong (the case where $\dot\gamma=1.1$ s\textsuperscript{-1}), the chain breaks. Computing the minimum of the force difference $F_\mathrm{d,x}+F_\mathrm{m,x}$, we reach the critical chain angle $\phi_\mathrm{c} \approx 33.5^\circ$ (plotted as a vertical dotted line). With the current method, however, the critical angle is slightly dependent on the applied shear rate. This is due to the excluded volume force feeding instantaneous distancing that arises from the viscous effects, thus inhibiting pure rotation and contributing to premature chain break. As can be seen from the plots of the decoupled magnetic attraction, the excluded-volume force reduces the stability of the chain. Still, it also delays the critical angle, as the minimum of the function $F_\mathrm{d,x}+F_\mathrm{a,x}$ is found slightly earlier, at $\phi_\mathrm{c}\approx28.3^\circ$.

Having determined the critical shear rates, we can define a strain rate range to test in oscillatory flow for either chain length. We aimed to test one strain rate for stable chains ($\dot\gamma_0<\dot\gamma_\mathrm{sub.}$), two within the intermediate range ($\dot\gamma_\mathrm{sub.}<\dot\gamma_0<\dot\gamma_\mathrm{brk.}$), and two past the breakdown threshold ($\dot\gamma_0>\dot\gamma_\mathrm{brk.}$ and $\dot\gamma_0\gg\dot\gamma_\mathrm{brk.}$). As such, for the chains with $l_\mathrm{chain} = 4$ and 10, we defined, respectively: $\dot\gamma_0\in[0.1, 0.3, 0.7, 1,5]$ s\textsuperscript{-1} and $\dot\gamma_0\in[0.01, 0.1, 0.7, 1, 5]$ s\textsuperscript{-1}. Ten strain values were tested between $0.1\leq\gamma_0\leq1$, running for 3 cycles ($t = 3\times2\pi/\omega$, sufficient to achieve a steady cycle) and the flow was driven by a top wall velocity $U=\dot\gamma_0\,h_\mathrm{c}\,\cos(\omega\,t)$, with $\omega = \dot\gamma_0/\gamma_0$. 

\begin{figure}[htp]
    \centering
    \small
    \includegraphics[width=\textwidth]{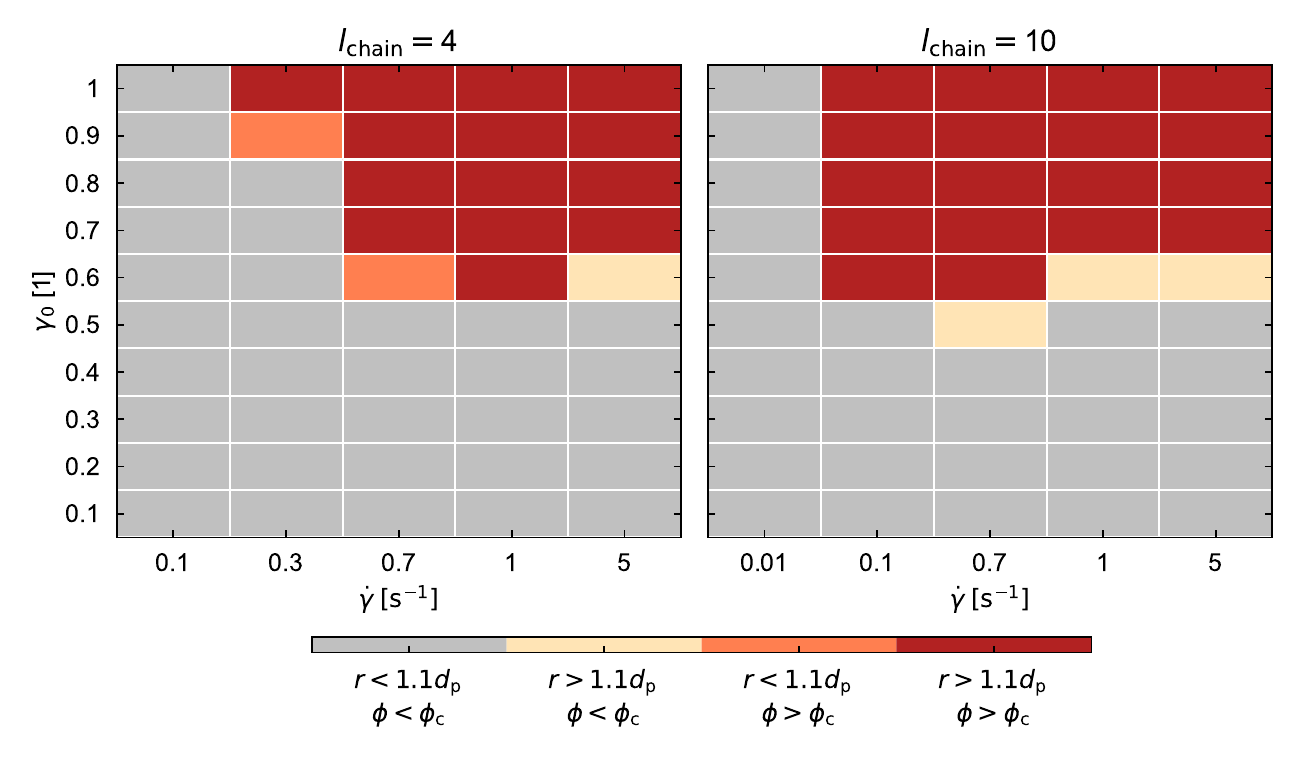}
    \caption{Passing conditions for chain break within the final simulated cycle of oscillatory shear. Chain is considered broken if $r>1.1\,d_\mathrm{p}$ and $\phi>\phi_\mathrm{c}$ (in red), and stable if $r<1.1d_\mathrm{p}$ and $\phi<\phi_\mathrm{c}$ (in gray).}
    \label{fig:Break_phase_space}
\end{figure}

Having the particle positions, we want to probe for chain subdivision. In steady shear, we could simply analyse the interparticle distance as post chain-break the particles continuously diverge, but in oscillatory shear, this method alone is insufficient due to the time-dependent nature of the flow and because the minimum distance for chain break is somewhat arbitrary. Here we choose to consider a possible subdivision if, within the final simulated cycle, the distance between the central particle-pair exceeds 10\% of their aggregated distance, $r/d_\mathrm{p}>1.1$, or if the critical angle is surpassed  $\phi>33.5^\circ$. In Figure~\ref{fig:Break_phase_space} are shown the passing conditions for chain break for $l_\mathrm{chain}=4$ and $10$.

As expected, for $\dot\gamma_0=0.1$ s\textsuperscript{-1} $<\dot\gamma_\mathrm{sub.}$ the chains never subdivide due to the magnetic dipole forces overpowering the weak viscous shear. At the other end of the strain rate range, $\dot\gamma_0=5$ s\textsuperscript{-1} $\gg\dot\gamma_\mathrm{brk.}$, for the majority of the cycle, the viscous shear dominates over the magnetic effects and the particles act as mere tracers. It could be expected that when the instantaneous rate decreases the dipole interactions would regain relevance and a degree of particle aggregation would be expected. However, because the maximum applied rate is so large, the time interval where the instantaneous rate is sufficiently decreased is very small, and the particles have no time to aggregate, even at small deformations where they remain relatively close. The resulting chain-break conditions for $\dot\gamma_0=5$ s\textsuperscript{-1} are therefore non-representative of the magnetic-chain dynamics, but more of the relative movement of individual particles. On the other hand, if the maximum strain rate is not so large, such as $\dot\gamma_0=1$ s\textsuperscript{-1} $>\dot\gamma_\mathrm{brk.}$, the time interval where the dipole forces are drowned out is small, being insufficient to provoke complete chain breakdown but sufficient for subdivision at large strains. On the other hand, for small $\gamma_0$, chain rotation is enough to comply with the imposed deformation and the chain remains unbroken. For $\dot\gamma_0=0.3$ and 0.7 s\textsuperscript{-1} the response is similar to $\dot\gamma=1$ s\textsuperscript{-1}, except that, similarly to complete chain breakdown, the chain will only subdivide if a sufficient portion of the cycle is above the critical rate. If not, the chain does not have enough time to rotate past the critical angle, which is the case of the smaller chain ($l_\mathrm{chain}=4$) for $\dot\gamma_0=0.3$ s\textsuperscript{-1}. 

Regarding the resulting conditions, we remind the reader that the selected critical distance for chain break is somewhat arbitrary and that the critical angle may depend on the simulation parameters. As such, for the central particle-pair, we plotted the instantaneous distance and chain angle on the final cycle (shown as supplementary material, Section \ref{sec:sup:Numerical_criteria}) and visually analysed the particle movement. Mostly, when both conditions are broken the chain clearly subdivides. Still, it is difficult to ascertain when only one critical quantity is surpassed, as the chains seem very near the subdivision threshold. In any case, the critical strain for chain subdivision seems to be between $0.5\lesssim\gamma_0\lesssim0.6$, which agrees with the magnetorheological change observed at $\gamma_0\approx0.5$. 

In some other works\citep{wang2021,li2004}, the beginning of viscous thinning may not agree with our critical strain, but this may be due to frequency dependence at larger $\omega$, larger particle concentrations that result in more complex microstructures or rheological differences of the continuous phase. The latter may be the case of the work of \citeauthor{wang2021}, who report $v_3$ sign shifts occurring much earlier ($\gamma_0\sim 0.01$), but their continuous phase is deeply non-Newtonian, which should alter the microstructural dynamics as the viscous forces diverge significantly from the Newtonian case. In any case, this hypothesis should be corroborated with other similar samples before drawing definitive conclusions.

\subsection{Magnetorheological conditions}

\begin{figure}[htp]
    \centering
    \small
    \includegraphics[width=\textwidth]{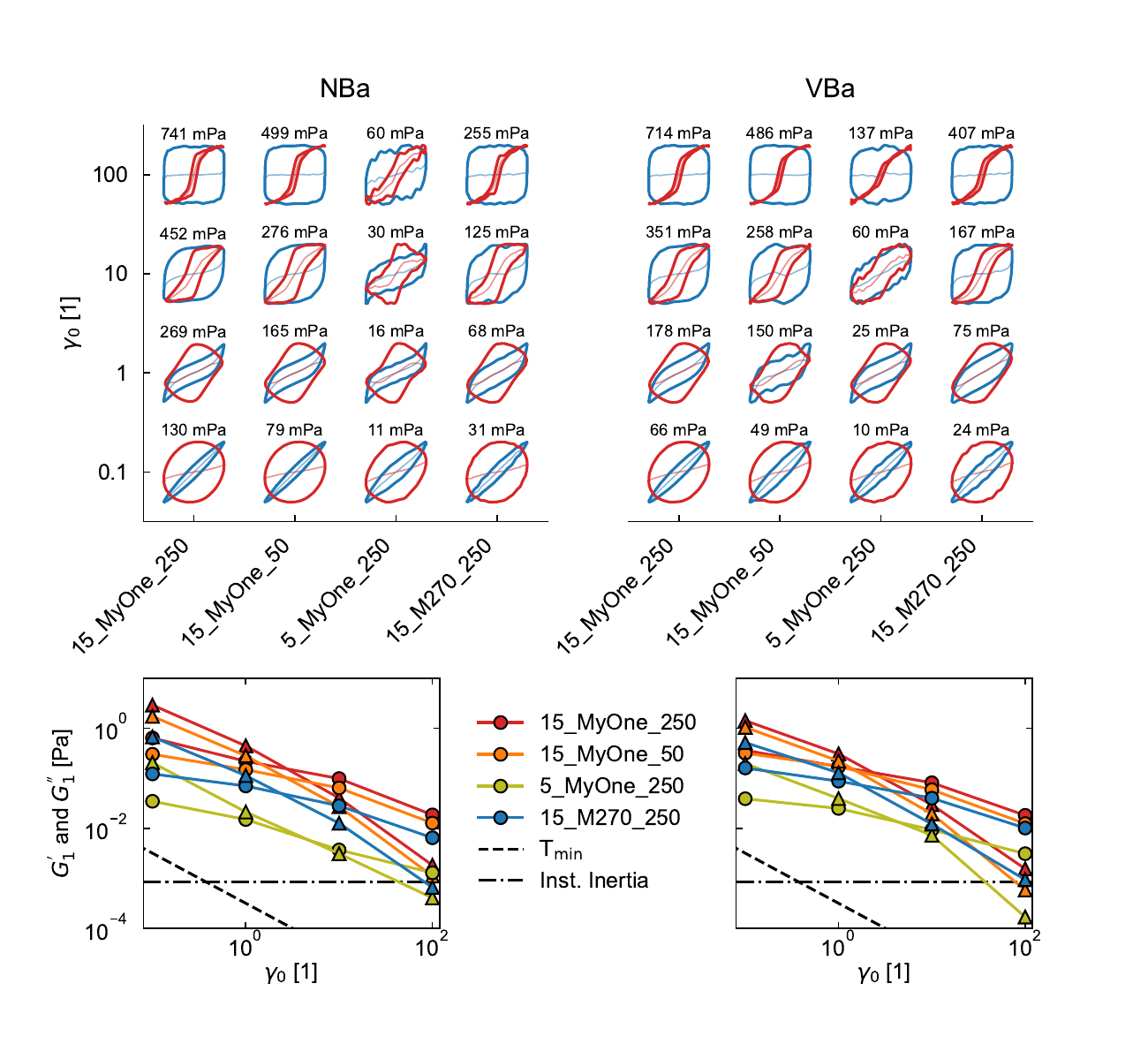}
    \caption{Oscillatory shear data gathered with the NBa (left) and VBa (right) samples (see Table \ref{tab:working_fluids}) at $\omega=0.1$ rad/s. (Top) raw Lissajous curves: blue and red curves represent, respectively, the elastic (stress vs strain) and viscous (stress vs strain rate) Lissajous curves. The elastic and viscous stresses are also lightly plotted in the respective colours and the maximum measured stress is presented above each plot. (Bottom) First-harmonic loss ($G''_1$, in circular markers) and storage ($G'_1$, in triangular markers) moduli and experimental limits associated with low-torque issues and instrument inertia (corrected for the geometry's gap-error).}
    \label{fig:Comparative_Lissajous}
\end{figure}

Returning to our experimental campaign, the magnetic field density, particle concentration and particle type were altered according to the protocol described in Table~\ref{tab:working_fluids}, with either analogue as the continuous phase. Figure~\ref{fig:Comparative_Lissajous} shows the raw Lissajous curves and first-harmonic moduli for the minimum tested frequency ($\omega=0.1$ rad/s). All measurements returned a weaker magnetorheological response, as was expected. This, however, led to error-ridden data at high frequency, which is why, here, we focus solely on the results gathered at $\omega=0.1$ rad/s. However, all the responses are given as supplementary material (\ref{sec:sup:comparisons}).

Again, no noticeable differences were introduced by the carrier fluids. Moreover, the changes to the magnetic field, particle concentration and particle type do not alter the non-linear behaviour displayed by the Lissajous curves, they do, however reduce the magnetorheological strength. Compared to the benchmark measurements, all alterations lead to a general decrease of both the storage and loss moduli. The least severe magnetorheological alteration is due to the field reduction to 50 mT, followed by the switch to the M270 particles, and most notoriously, the particle concentration reduction to 5 wt\%.  This weakening of the magnetorheological response leads to an approach of the experimental limits as the samples soften at high strain. Particularly, the 5 wt\% measurements display clear oscillations of the Lissajous curves, meaning the current setup is not suitable for the magnetorheological characterisation of weak MR fluids. In any case, these results agree with the steady shear data, and the reasoning for the magnetorheological hierarchy between our measurements has already been discussed.

As mentioned above, the alterations did not affect the non-linear response evaluated at the tested strain values. This is further corroborated by the FT-Chebyshev analysis. For the sake of conciseness, the data given by MITLaos is not shown here but is given as supplementary material (\ref{sec:sup:comparisons}). The relevant moduli ($G'_M$ and $G'_L$) and dynamic viscosities ($\eta'_M$ and $\eta'_L$) all show the same tendency as the benchmark measurement (displayed in Figure~\ref{fig:15wt_MyOne_250mT_MITlaos}), diminishing with strain (and strain rate) increase. $G'_M$ and $G'_L$ diverge past $\gamma_0>0.1$, with $G'_M$ decreasing more acutely. For $\dot\gamma_0=0.1$ s\textsuperscript{-1}, $\eta'_M<\eta'_L$ except for the 5wt\% measurement where $\eta'_M\lesssim\eta'_L$, which is probably due to low-torque errors. In any case, for all measurements, $\eta'_L$ decreases faster than $\eta'_M$. The third-order Chebyshev coefficients of the comparative measurements also display the same behaviour as the benchmark case. The elastic coefficient, $e_3$, always grows positive from zero at $\gamma_0=0.1$ (except for the 5 wt\% measurement, where $e_3/e_1\approx5\%$, probably from experimental error). The viscous counterpart, $v_3$, always crosses over from positive to negative between $0.1<\dot\gamma_0<1$, which may imply that the rheological and microstructural discussion is also valid for each of the alterations tested. Nonetheless, as we have seen, this sparse strain data set may hide important behaviours, and we believe the alterations tested should provoke some behavioural differences, as for example in the strain overshoot of $G''_1$.

\section{Conclusions}
\label{sec:conclusions}

Steady and oscillatory shear measurements were conducted using two blood analogues, one Newtonian and one viscoelastic, both seeded with magnetic particles and subjected to an external, uniform magnetic field perpendicular to the flow. The influence of magnetic field density, particle concentration and particle type was also evaluated by independently altering each parameter from a defined benchmark: 15wt\% of MyOne particles in a 250 mT magnetic field. The magnetic field was reduced to 50 mT, the particle concentration to 5 wt\%, and the MyOne particles were replaced with the M270.

Under steady shear, the benchmark case returned a massive increase in low-shear viscosity and a clear shear-thinning behaviour, as the magnetically-induced microstructure progressively breaks down due to competing viscous shearing forces gaining relevance as the shear rate increases. The response is well described by the Casson model, even though the yield stress could not be directly observable in the flow curves, possibly due to apparent slip on the bottom plate of the rheometer. In oscillatory shear, the initial measurements displayed a practically frequency-independent response; however, having $\omega\gtrsim1$ rad/s gave rise to critical errors likely due to inertial effects. A transition set was observed with increasing strain: linear viscoelasticity to non-linear viscoelasticity to viscoplasticity, typical of field-active fluids\citep{parthasarathy1999,li2004}. The FT-Chebyshev analysis revealed that the elastic softening was driven by the strain rate, whereas the viscous non-linearity was more complex. A finer amplitude sweep revealed a type III behaviour\cite{hyun2002}, with a weak overshoot of $G_1''$ at the beginning of the non-linear regime associated with small structural rearrangements. The initial increase in $G_1''$ was accompanied by a positive contribution of the third-order viscous Chebyshev coefficient, $v_3>0$, which relates to viscous thickening driven by the strain rate. Increasing the strain, when $G_1''$ begins its descent, $v_3$ momentarily remained positive, which may be related to a shift of the driving mechanism of the viscous thinning non-linearity from the strain to the strain rate (at $\gamma_0\approx0.5$).

A magnetised-particle chain was modelled under oscillatory flow, and a critical strain was found between $0.5\lesssim\gamma_0\lesssim0.6$, above which chain rotation alone no longer complied with the imposed deformation and the chain was forced to subdivide. As such, the numerical results corroborate the experimental observation, but more experimental data with similar magnetorheological fluids are necessary to confirm the hypothesis. 

All the alterations to the benchmark measurement resulted in a weaker magnetorheological response. In steady shear, the low-shear viscosity was reduced, particularly for the lesser particle concentration, and some evidence of apparent slip of the dispersed phase was noticed, meaning in future measurements both the geometry and the bottom plate must have significant surface roughness to avoid slip. In any case, the experimental data generally collapsed onto a single master curve dependent on the Mason number. In oscillatory shear, the alterations showed no significant differences in the non-linear behaviour, but finer sweeps are required to confirm this. The weakened magnetorheological response led to significant errors, which could compromise measurements with whole blood. The increase in the geometry radius is the best solution to widen the experimental window, but, for now, the geometrical restrictions associated with commercially available magnetorheological cells may be impeding a magnetorheological characterisation of blood under oscillatory shear.

\pagebreak

\section*{Acknowledgements}

This work was financially supported by national funds through the FCT/MCTES (PIDDAC), under the project PTDC/EME-APL/3805/2021 (DOI 10.54499/PTDC/EME-APL/3805/2021), by the FEDER funds through COMPETE2030 and with financial support of FCT, I.P., within the framework of the project PROMisH - Potential of Rheologically-Optimized Magnetic Emulsions in Hemotherapy, with nr. 16207, operation code at the Funds Platform COMPETE2030-FEDER-00736400 and FCT code 2023.17217.ICDT With DOI https://doi.org/10.54499/2023.17217.ICDT., LA/P/0045/2020, UIDB/00532/2020 and UIDP/00532/2020, and the program Stimulus of Scientific Employment, Individual Support-2020.03203.CEECIND. R.R. acknowledges FCT for funding support under scholarship No. 2025.03368.BD.





\bibliographystyle{unsrtnat}
\bibliography{bibliography.bib}


\newpage
\pagestyle{plain}

\renewcommand{\thepage}{S\arabic{page}}
\renewcommand{\thesubsection}{S\arabic{subsection}}
\renewcommand{\thetable}{S\arabic{table}}
\renewcommand{\thefigure}{S\arabic{figure}}
\renewcommand{\theequation}{S\arabic{equation}}

\renewcommand\bibsection{\subsection*{\refname}}

\setcounter{page}{1}
\setcounter{subsection}{0}
\setcounter{table}{0}
\setcounter{figure}{0}
\setcounter{equation}{0}

\nolinenumbers
\section*{Supplementary material}
\addcontentsline{toc}{section}{Supporting information}
\label{sec:suplementary}
\subsection{Apparent slip with the CP20 MRD}
\label{sec:sup:preliminary}
\begin{figure}[htp]
\centering
\includegraphics[width=\linewidth]{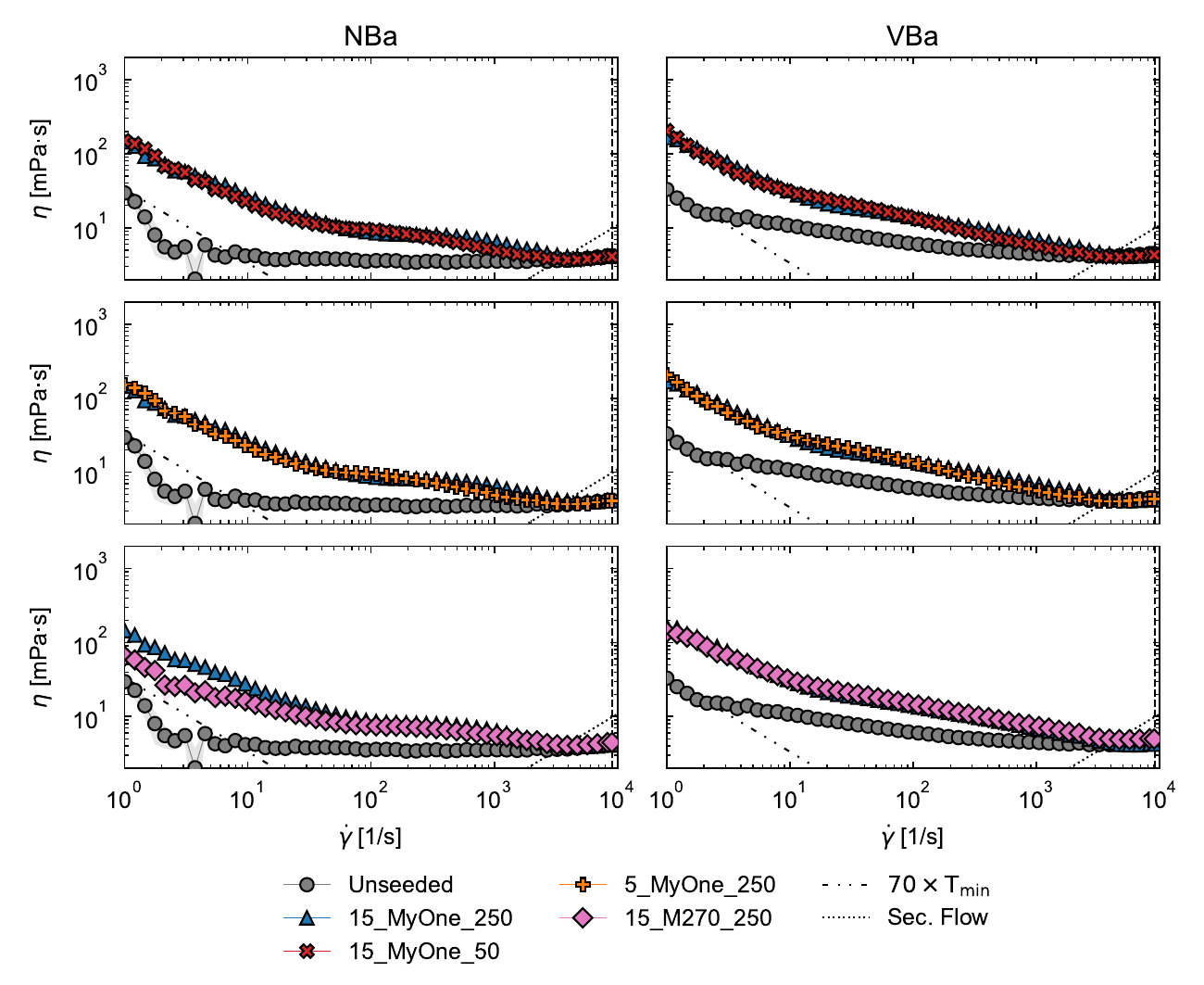}
\caption{Viscosity curves obtained with the NBa (left column) and VBa (right column), seeded with MyOne particles at 15 wt\% under the influence of 50 and 250 mT (top row), with MyOne particles at 5 and 15 wt\% under 250 mT (middle row), and with MyOne and M270 particles at 15 wt\% under 250 mT (bottom row). Adjusted low-torque and secondary flow limits are also shown in each graph. Data gathered with the CP20 MRD geometry.}
\label{fig:CP_flow_curves}
\end{figure}

Following the experimental campaign described at the end of Section~\ref{sec:methods}, in Figure~\ref{fig:CP_flow_curves} are shown the flow curves, obtained with the CP20 MRD, of the magnetorheological benchmark measurements (15 wt\% of MyOne particles under a magnetic field of 250 mT) compared to the relevant parameter alterations (magnetic field density, particle concentration and particle type) and the unseeded blood analogues.

The introduction of the magnetised particles returns a rheological response that appears practically independent of the applied field, particle concentration, and particle type. In all cases, there is an increase in the low-shear viscosity when compared to the unseeded analogues. Increasing the shear rate leads to a viscosity reduction, tending to that of the continuous phases, but not resembling the usual, smooth shear-thinning behaviour. At middling shear rates, there seems to be a tendency for a pseudo-Newtonian plateau that is more pronounced for the samples with the NBa as the continuous phase. This response at middling shear rates is reminiscent of previous reports of particle depletion near the smooth geometry walls. Because the continuous phase is less viscous, a substantial part of the imposed deformation is applied in this region, returning a lesser measured viscosity than if the particles were perfectly dispersed\citep{barnes1995,vicente2004}. One way to eliminate this apparent particle slip is to employ geometries with enhanced surface roughness, as the serrated parallel-plate geometry for magnetorheological measurements, the PP20 MRD P2. 

\begin{figure}[htp]
\centering
\begin{minipage}{\textwidth}
\centering
Unseeded
\end{minipage}
\begin{minipage}{\textwidth}
\centering
\includegraphics[width=\linewidth]{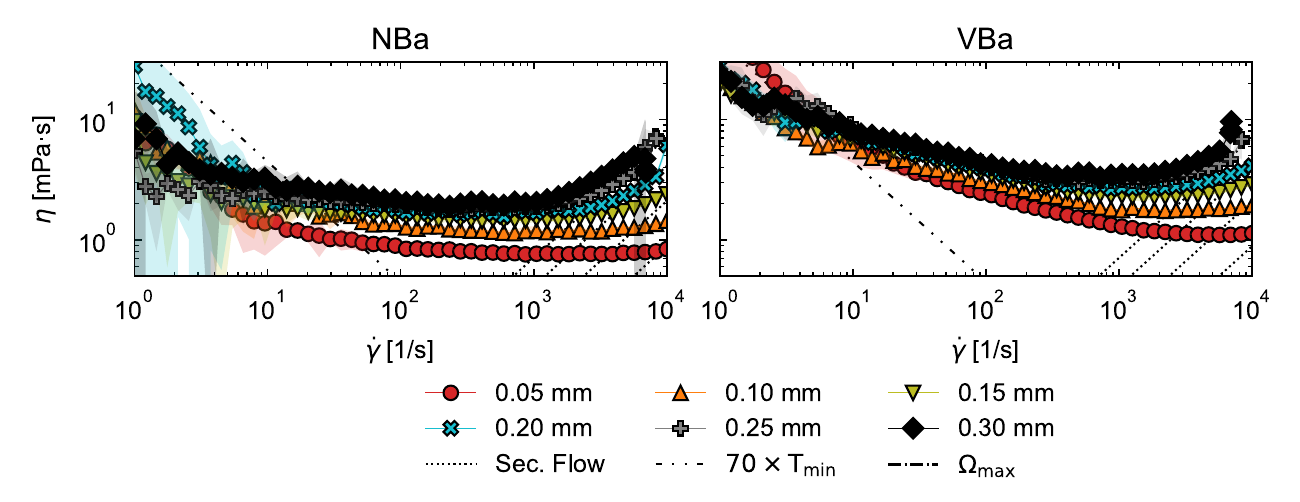}
\end{minipage}
\caption{Flow curves obtained with the (left) NBa and (right) VBa, without magnetic particles (0 wt\%), with the PP20 MRD P2 geometry at different commanded gap heights ($0.05 \leq h_\mathrm{c} \leq 0.30$ mm). Experimental limits associated with low-torque issues, maximum rotational velocity and secondary flows are also presented.}
\label{fig:PP20_MRD_P2_Std_Shear_No_particles}
\end{figure}

First, measurements were conducted with either fluid without magnetic particles and no magnetic field application to evaluate the suitability of the geometry. The measurements were conducted at different commanded gap heights, $0.05 \leq h_\mathrm{c} \leq 0.30$ mm, and the resulting flow curves are shown in Figure~\ref{fig:PP20_MRD_P2_Std_Shear_No_particles}.

\begin{figure}[htp]
\centering
\begin{minipage}{\textwidth}
\centering
Unseeded
\end{minipage}
\begin{minipage}{\textwidth}
\centering
\includegraphics[width=\linewidth]{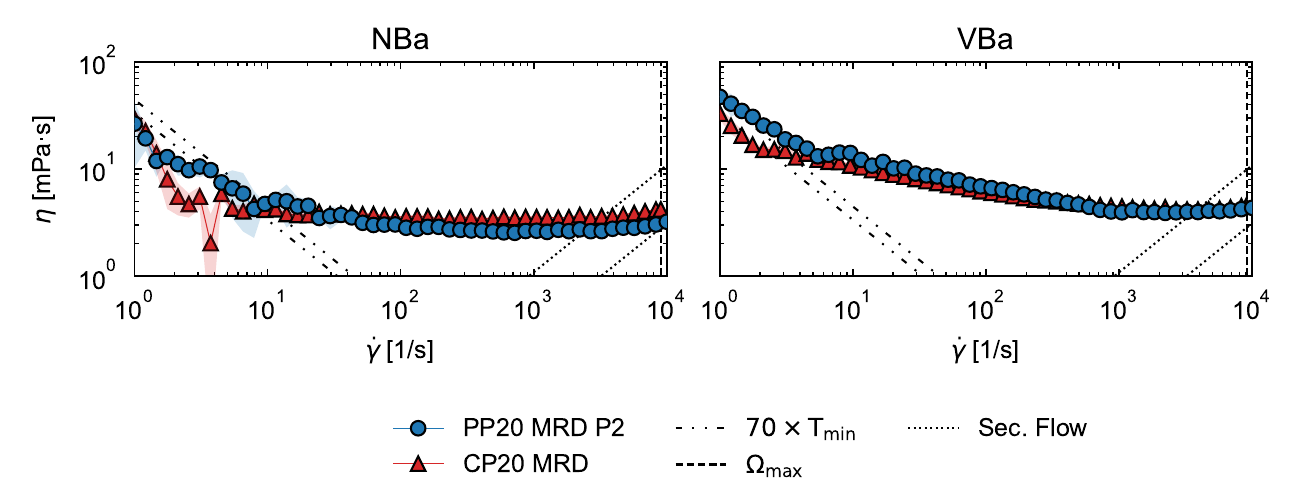}
\end{minipage}
\caption{Flow curves obtained with the (left) NBa and (right) VBa, with no magnetic particles (0 wt\%) nor field applied, with the PP20 MRD P2 ($h_\mathrm{c} = 0.10$ mm) and CP20 MRD geometries. Experimental limits associated with low-torque issues, maximum rotational velocity and secondary flows are also presented (corrected for the gap error of the PP20 MRD P2).}
\label{fig:PP20_MRD_P2_Std_Shear_No_particles_corrected}
\end{figure}

The flow curves clearly display a measured-viscosity dependence on the commanded gap height, yielding lower viscosities as the gap decreases. This can be explained either by slip effects or a gap-error\citep{vleminckx2016}, but the latter seems more likely as detailed or roughened geometries allow the sample to flow within the solid structures\citep{carotenuto2013,nickerson2005}. Having measured-viscosity data, $\eta_\mathrm{m}$, at different commanded gaps, $h_\mathrm{c}$, allows for an estimation of both the gap error, $\epsilon$, and the samples' true viscosities, $\eta_\mathrm{r}$, by fitting the expression adapted from \citeauthor{kramer1987}:
\begin{equation}
    \frac{h_\mathrm{c}}{\eta_\mathrm{m}} = \left(\frac{1}{\eta_\mathrm{r}}\right)h_\mathrm{c} + \frac{\epsilon}{\eta_\mathrm{r}}\,,
    \label{eq:gap error estimate}
\end{equation}
to the data at a specific shear rate. As can be seen, particularly for larger gaps, there are significant fluctuations in the flow curves. This means that there will also be variability in the gap-error and true viscosity estimation. Therefore, we chose to fit the expression to data gathered at multiple shear rates within a useful range ($10^2\leq\dot\gamma\leq 10^3$ s\textsuperscript{-1}), given that at low and high shear the data is affected by experimental limitations, particularly large-gap data\citep{rodrigues2025}. The gap error was estimated as $\epsilon = 123$ \textmu m (standard error of 13 \textmu m) and the flow curves obtained at a commanded gap of $h_\mathrm{c} = 0.10$ mm are shown, corrected, in Figure~\ref{fig:PP20_MRD_P2_Std_Shear_No_particles_corrected}. To compare the results of the PP20 MRD P2, in Figure~\ref{fig:PP20_MRD_P2_Std_Shear_No_particles_corrected} flow curves obtained with the CP20 MRD are also shown. The secondary flow and maximum rotational velocity limits are also gap-dependent and, because the gap-error is quite large, these were also corrected to the estimated true gap of the PP20 MRD P2, $h = h_\mathrm{c}+\varepsilon = 223$ \textmu m. The gap-error estimate seems to reasonably correct the PP20 MRD P2 results, which are now similar to the data gathered with the CP20 MRD.

\begin{figure}[htp]
\centering
\begin{minipage}{\textwidth}
\centering
15\_MyOne\_250 (NBa)
\end{minipage}
\begin{minipage}{\textwidth}
\centering
\includegraphics[width=0.7\linewidth]{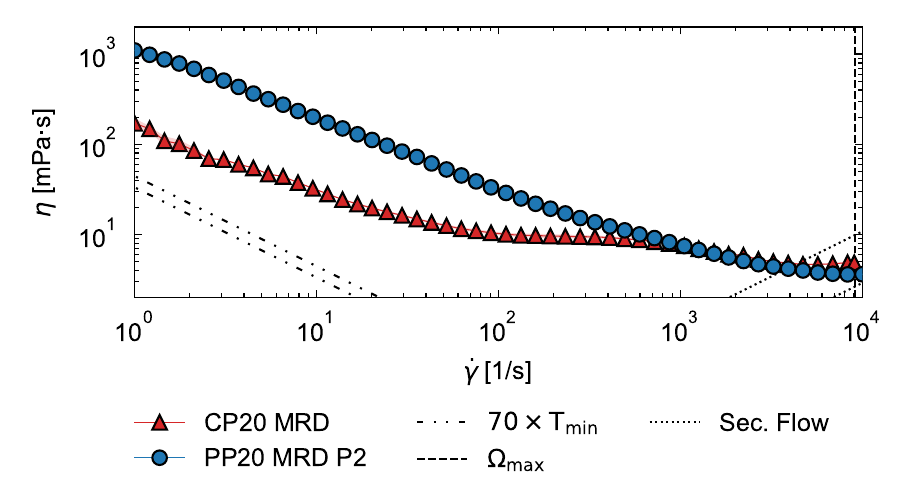}
\end{minipage}
\caption{Flow curves obtained with the NBa sample with 15 wt\% of MyOne particles, on the PP20 MRD P2 ($h_\mathrm{c} = 0.10$ mm) and CP20 MRD geometries with an applied field density of $B = 250$ mT. Experimental limits associated with low-torque issues, maximum rotational velocity and secondary flows are also presented (corrected for the gap error of the PP20 MRD P2).}
\label{fig:PP20_MRD_P2_Std_Shear_Magnetic_corrected}
\end{figure}

Regarding the apparent slip in magnetorheological measurements, in Figure~\ref{fig:PP20_MRD_P2_Std_Shear_Magnetic_corrected} are shown the flow curves of the NBa benchmark sample (15 wt\% of MyOne particles under a magnetic field of 250 mT) gathered with the CP20 MRD and the PP20 MRD P2 (at a commanded gap of $h_\mathrm{c} = 0.10$ mm and corrected for the gap-error). The PP20 MRD P2 flow curve shows a smooth shear-thinning response of the benchmark sample, with no evidence of the previously-noted pseudo-Newtonian plateau and with a much higher low-shear viscosity than with the CP20 MRD (almost one order of magnitude). Indeed, it seems that the serrated geometry was able to gather data free from the apparent slip effect, returning the expected response. As such, we can assume the effects of apparent slip have been minimised; however, because our bottom plate is not detailed, nor was it treated with any surface finishing to enhance the roughness, we must beware, particularly because testing the gap variation seems not to be able to predict apparent slip in magnetorheological tests\citep{jonkkari2012}.

Looking back at Figures~\ref{fig:PP20_MRD_P2_Std_Shear_No_particles}, \ref{fig:PP20_MRD_P2_Std_Shear_No_particles_corrected} and \ref{fig:PP20_MRD_P2_Std_Shear_Magnetic_corrected}, the experimental window is limited, in the present case, by three issues. At low shear rates, the rheometer's minimum torque ($\mathrm{T}_\mathrm{min} = 1$ nN$\cdot$m) limits the accurately-measured viscosity, but surface tension forces acting along an asymmetrical contact line can generate a residual torque orders of magnitude larger than the rheometer's minimum torque\citep{johnston2013}. From an analysis of the data, multiplying the minimum-torque limit by a factor of 70 seems to reasonably predict the large uncertainty in low-shear data. At high shear, sample inertia becomes an issue and radial velocity components result in an apparent shear-thickening, which is predicted by the secondary-flow limit\citep{turian1972}. Regarding the non-corrected flow curves of Figure~\ref{fig:PP20_MRD_P2_Std_Shear_No_particles}, the apparent shear-thickening onsets much earlier than expected from the secondary flow limit. This should be due to the gap-error, as larger gaps are more susceptible to secondary flow onset. Finally, the rheometer's maximum rotational velocity ($\Omega_\mathrm{max} = 314$ rad/s) limits the achievable shear rate, but this will not affect the results of the PP20 MRD P2 because we are employing a relatively small gap, $h_\mathrm{c} = 0.10$ mm. 

\pagebreak
\subsection{Bingham model fit to the obtained flow curves}
\label{sec:sup:Bingham}

\begin{figure}[htp]
\centering
\includegraphics[width=\linewidth]{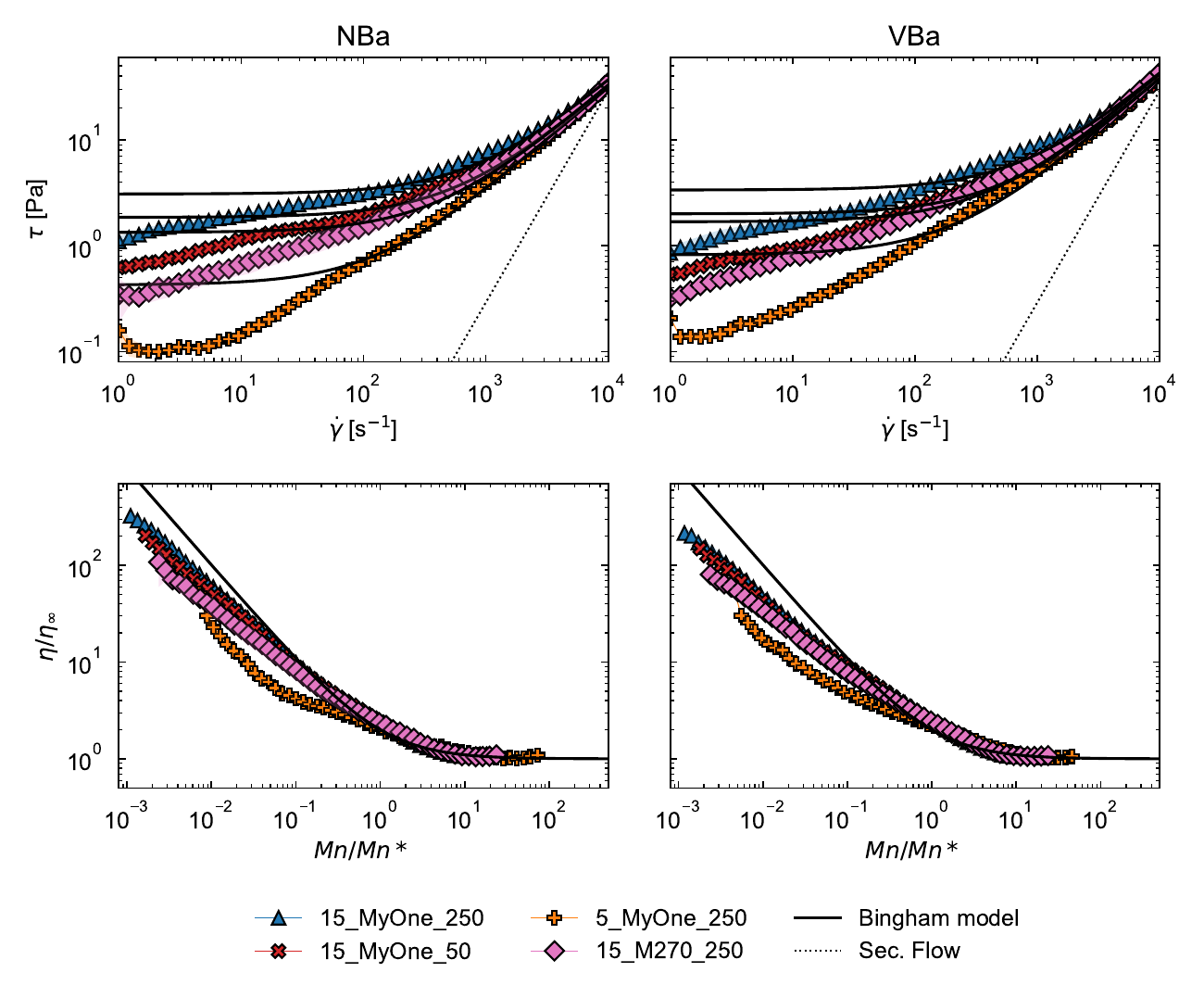}
\caption{Flow curves with Bingham model fits (top row) and dimensionless viscosity as a function of the reduced Mason number (bottom row), for each tested particle type, concentration and magnetic field density combination, for samples with the NBa (left) and VBa (right) as continuous phase (unseeded data not shown). Secondary flow limit is shown along with the flow curves (top row).}
\label{fig:Bingham_master_curves}
\end{figure}

The top row of Figure \ref{fig:Bingham_master_curves} shows the Bingham model fits to the experimental flow curves. The bottom row displays the dimensionless viscosity curves as a function of the reduced Mason number (Equations \ref{eq:Mason_number} and \ref{eq:reduced_Mason_number} of the main text), along with the Bingham model prediction:
\begin{equation}
    \tau = \tau_\gamma+\eta_\infty\dot\gamma \Leftrightarrow \frac{\eta}{\eta_\infty}=1+\left(\frac{Mn}{Mn^*}\right)^{-1}\,,
\end{equation}

\pagebreak
\subsection{Oscillatory shear measurements with PP50 geometry}
\label{sec:sup:PP50}

\begin{figure}[htp]
    \centering
    \begin{minipage}{\textwidth}
    \centering
    (PP50) Unseeded
    \end{minipage}
    \centering
    \begin{minipage}{\textwidth}
    \centering
    \small
    \includegraphics[trim={0cm 1cm 0cm 1.5cm},clip,width=\textwidth]{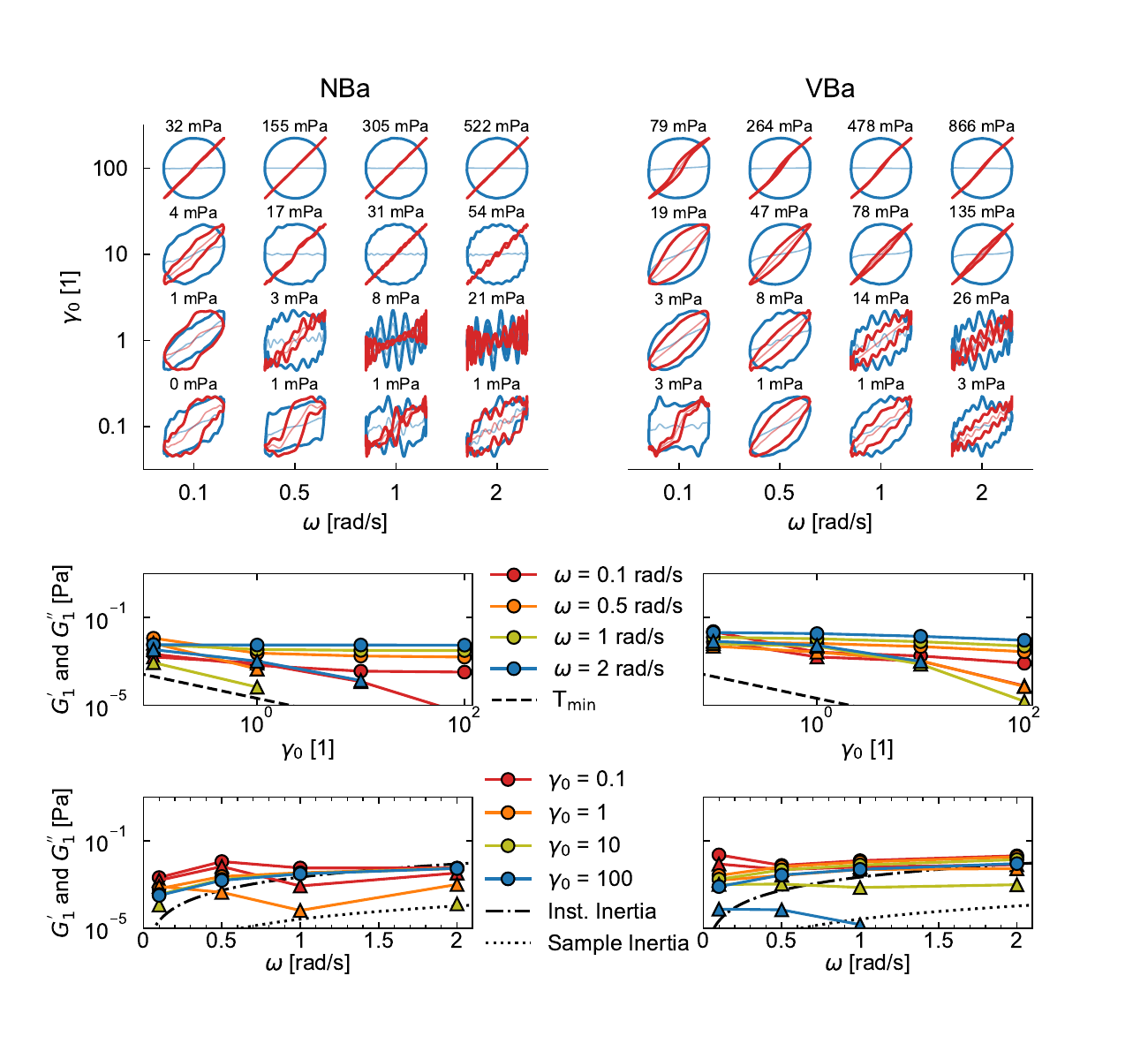}
    \end{minipage}
    \caption{Oscillatory shear data gathered with the unseeded (left) NBa and (right) VBa, with a standard, smooth PP50. (Top) Pipkin diagrams depicting the Lissajous curves. The blue and red curves represent, respectively, the elastic (stress vs strain) and viscous (stress vs strain rate) Lissajous curves. The elastic and viscous stresses are also plotted lightly in their respective colours, and the maximum measured stress is presented above each plot. (Bottom) First-harmonic loss ($G''_1$, in circular markers) and storage ($G'_1$, in triangular markers) moduli and experimental limits associated with low-torque issues and instrument and sample inertia.}
    \label{fig:Oscillatory_PP50_0wt_0mT}
\end{figure}

The same oscillatory shear measurements were conducted with the  PP50 measuring geometry ($I\approx0.0136$ mN$\cdot$m$\cdot$s\textsuperscript{2}, $h_\mathrm{c}=0.1$ mm). Figure~\ref{fig:Oscillatory_PP50_0wt_0mT} shows the obtained Pipkin diagrams and moduli plots.

Analysing the Pipkin diagrams, we can now see the expected fluid responses at high deformation ($\gamma_0 = 100$). The NBa returned practically perfectly circular elastic Lissajous curves, and the viscous curve was the expected positive-diagonal. Moreover, the elastic and viscous stress plots also show practically straight lines, indicating the purely viscous linear response we expect from a Newtonian fluid. On the other hand, at this imposed strain, we can see the VBa presenting a slight bending of the viscous Lissajous, an indicator of additional harmonics emerging in the NLVE. Reducing the deformation to $\gamma_0 = 10$, the NBa shows some apparent viscoelastic behaviour at small frequencies, which, looking at the moduli plots, could be due to surface-tension torque. As such, this could also be the cause for the VBa's viscoelastic response (at $\gamma_0 = 10$ and $\omega<1$ rad/s) despite its moduli being slightly larger. Nevertheless, at this amplitude, increasing the frequency still somewhat returns the NBa's purely viscous behaviour (despite some oscillations possibly due to inertial issues), while the VBa presents a slight viscoelasticity which agrees with its very subtle elastic nature ($\lambda_\mathrm{CaBER} = 0.35$ ms), and there is observable bending of the elastic stress which indicates we are still obtaining a non-linear response. Considering lower strains, $\gamma_0<10$, because the NBa data is no longer representative of a purely viscous behaviour, we are wary of gathering conclusions of its viscoelastic counterpart. As such, we assume the data for $\gamma_0<10$ is no longer representative of either fluid's rheology. 

As discussed, there is evidence of surface tension torque at low frequency up to middling amplitudes (a non-negligible apparent elasticity of the NBa). Looking at the low-torque limit, we can presume, thus, that surface tension has indeed shifted the limit considerably and the data gathered at low frequency and amplitude should be viewed with scepticism. Considering inertial effects, it is challenging to attribute the fault to instrument or sample inertia; however, errors still dominate at high frequencies and small amplitudes. Despite these issues, the increase in diameter, possibly aided by the elimination of the gap-error (the estimated PP50 gap-error is practically insignificant\citep{rodrigues2025}, $\varepsilon\approx3$ \textmu m), successfully widened the experimental window, allowing us to visualise the analogues' response to oscillatory shear. In the current state, this is useful for standard measurements with samples that exhibit subtle rheological properties.

\pagebreak
\subsection{Numerical analysis: spatial discretization}
\label{sec:sup:mesh}

The calculation of the particle dynamics is not directly dependent on the employed mesh, as the only acting forces are inter-particle (dependent on the uniform magnetic field and their own positions) and viscous effects. As such, to effectively model the chain behaviour, the mesh selection criteria should be the quality of the resulting flow field. Our flow, however, is very simple, where a single moving boundary generates a linear velocity profile. Indeed, such a flow field could be effectively modelled by single 1\textsuperscript{st}-order elements along the channel/gap height. However, the lateral open boundaries act as inlets/outlets to large reservoirs where curved streamlines are expected, which ought to require a finer mesh. 

\begin{figure}[htp]
    \centering
    \small
    \includegraphics[width=\textwidth]{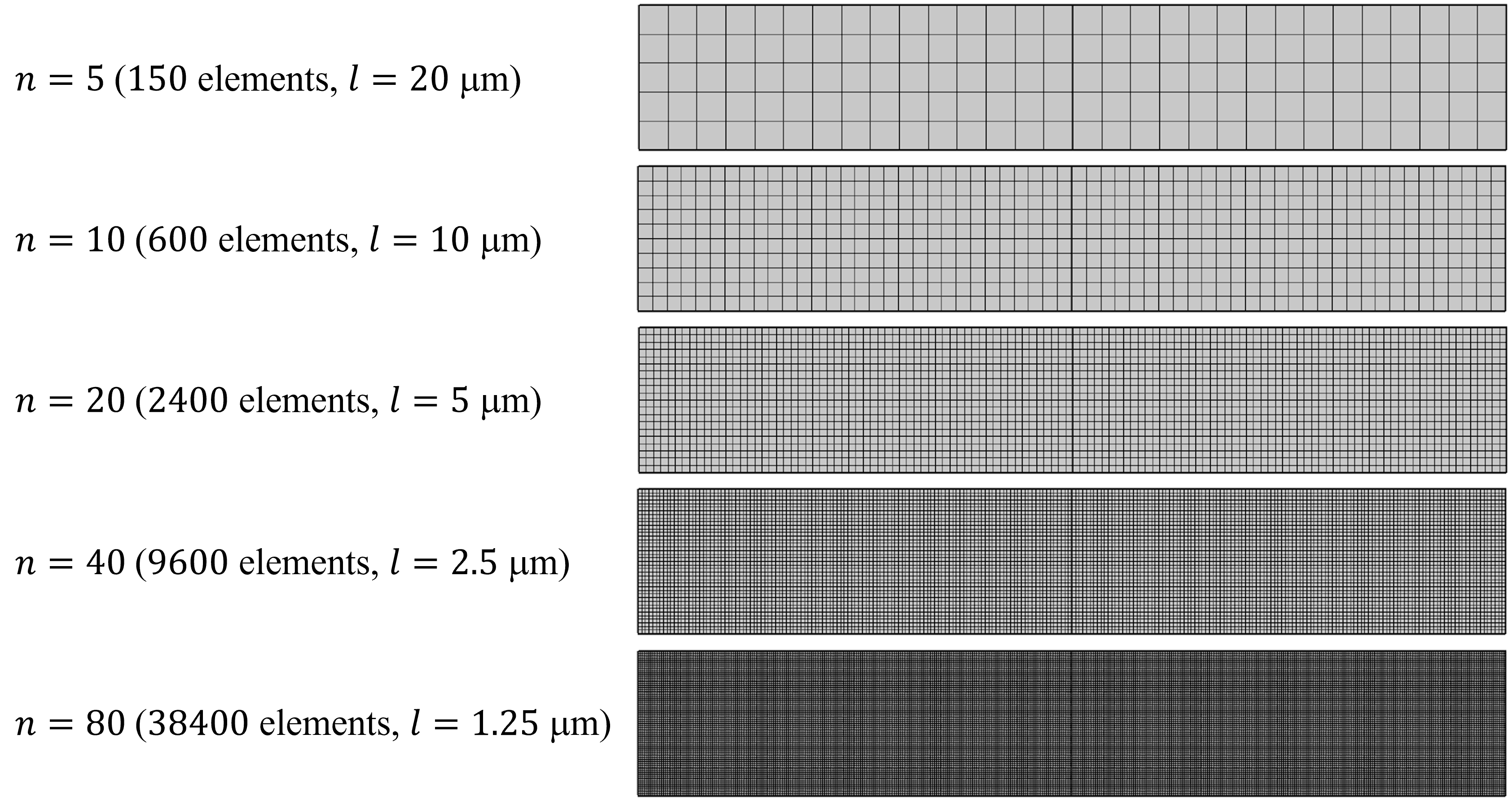}
    \caption{Tested meshes, identified by the number of elements along the channel/gap height, with respective number of elements and element size}
    \label{fig:meshes}
\end{figure}

We chose to mesh the domain with quadrilateral elements of equal size, which was defined by partitioning the channel/gap height into $n$ segments: $l=h_\mathrm{c}/n$, meaning the whole mesh was composed of $6\times n^2$ elements. We tested 5 meshes as shown in Figure \ref{fig:meshes}.

\begin{figure}[htp]
    \centering
    \small
    \includegraphics[width=\textwidth]{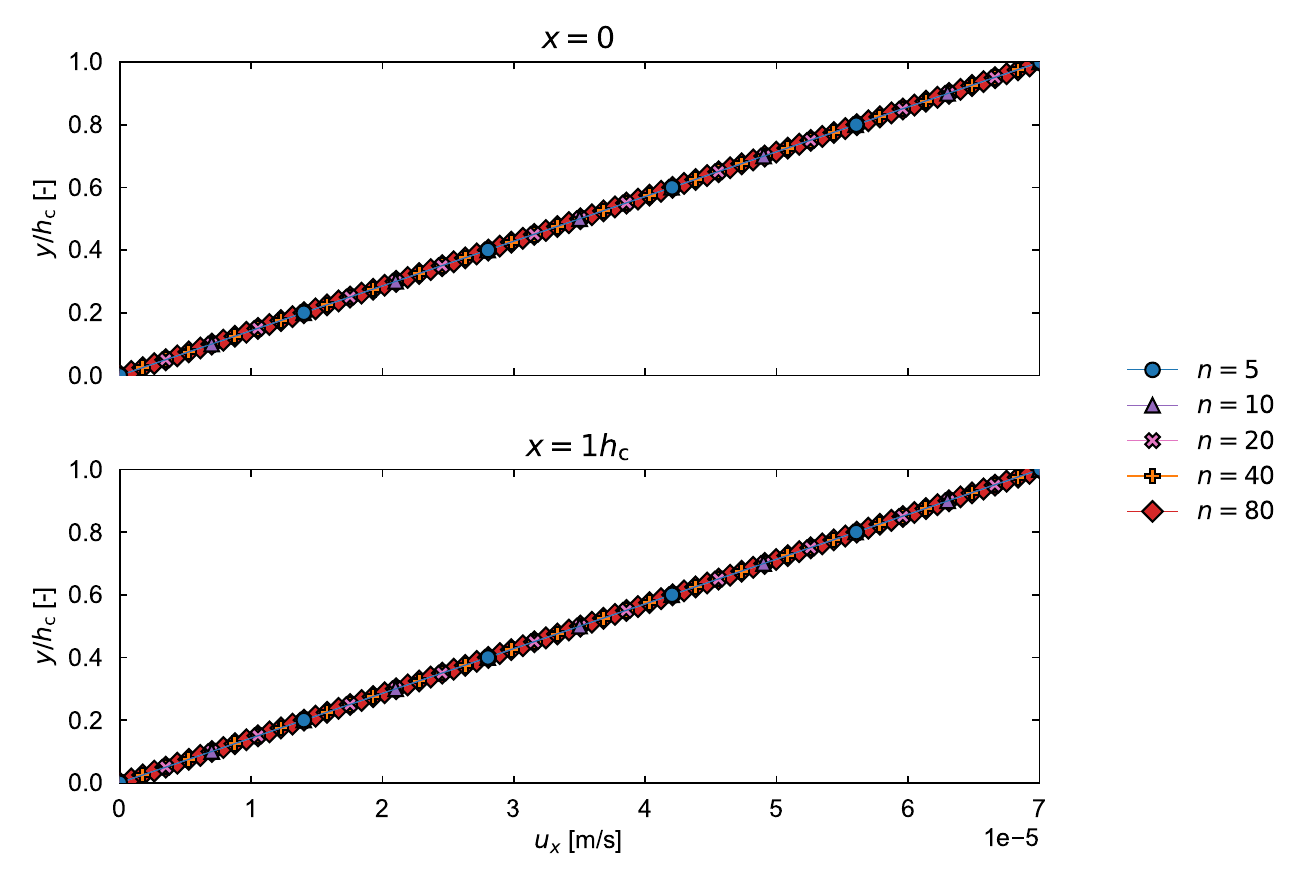}
    \caption{Axial-velocity profiles, $u_x$, at $x=0$ and $x=h_\mathrm{c}$, for each tested mesh (identified by the number of elements along the channel/gap height, $n$).Data taken at the end of the final oscillatory cycle, $t=3\,(2\pi/\omega)$ ($\gamma_0=1$ and $\dot\gamma_0=0.7$ s\textsuperscript{-1}).}
    \label{fig:mesh_vertical}
\end{figure}

First, we evaluated the resulting flows in the region of interest, $-h_\mathrm{c}\leq x\leq h_\mathrm{c}$ (dependent on the selected strain amplitude, $\gamma_0=1$, which will be discussed further on), by plotting the axial-velocity profiles, $u_x$, at two locations: $x=0$ and $x = h_\mathrm{c}$ (see Figure \ref{fig:Numerical_model_schematic} and its caption for information on the coordinate system). These profiles, plotted in Figure \ref{fig:mesh_vertical}, show that the flow appears to be independent of the selected mesh, with all profiles superimposing in a triangular shape, as expected.

\begin{figure}[htp]
    \centering
    \small
    \includegraphics[width=\textwidth]{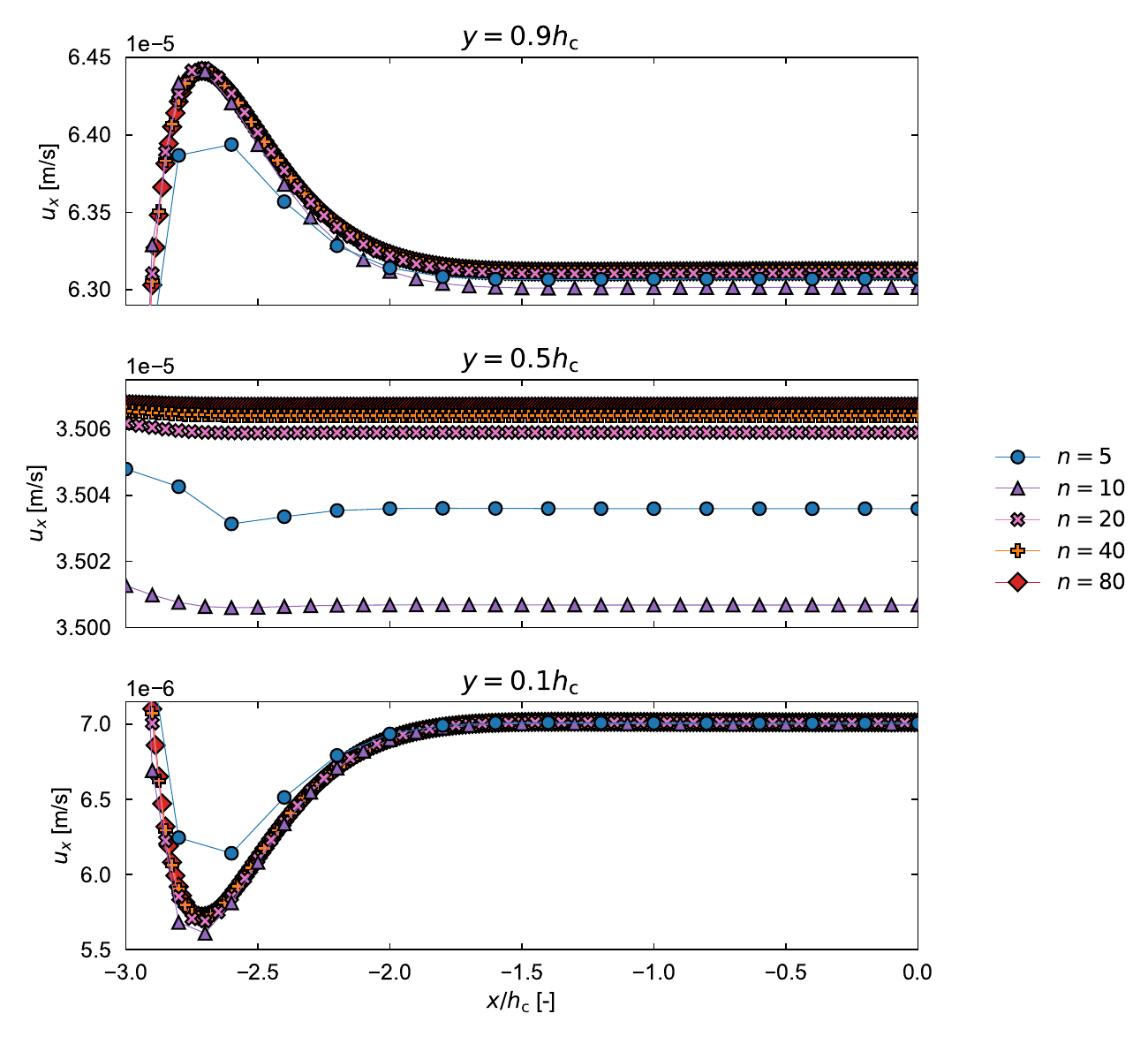}
    \caption{Axial-velocity profiles, $u_x$, at $y=0.1h_\mathrm{c}$, $0.5h_\mathrm{c}$ and $0.9h_\mathrm{c}$, for each tested mesh (identified by the number of elements along the channel/gap height, $n$). For clarity, only the first half of the channel is shown, $-3h_\mathrm{c}\leq x \leq 0$ (the flow is symmetric). Data taken at the end of the final oscillatory cycle, $t=3\,(2\pi/\omega)$ ($\gamma_0=1$ and $\dot\gamma_0=0.7$ s\textsuperscript{-1}).}
    \label{fig:mesh_horizontal}
\end{figure}

However, as mentioned, differences should appear near the open boundaries where rougher meshes may not be able to accurately represent inlet/outlet flow. Moreover, the velocity profiles along a gap-spanning plane may not be able to convey some subtle differences between the flow fields. To evaluate this, in Figure \ref{fig:mesh_horizontal} are shown velocity profiles along the channel at three heights: $y=0.1h_\mathrm{c}$, $0.5h_\mathrm{c}$ and $0.9h_\mathrm{c}$. Indeed, there are discrepancies between the tested meshes, with the roughest mesh, $n=5$, struggling to model the inlet/outlet flow, and increasing $n$ above 20 does not return significant improvements. The entrance effects seem to become negligible at $-1.5 h_\mathrm{c}\gtrsim x \gtrsim 1.5 h_\mathrm{c}$, meaning the region where the particles will travel should not be affected. Nevertheless, we opted for the $n=20$ mesh, as it seems to yield the best cost/benefit, and it is also similar to other, previously employed configurations \citep{rodrigues2025}.

\pagebreak
\subsection{Numerical analysis: interparticle distance and chain angle}
\label{sec:sup:Numerical_criteria}
\begin{figure}[htp]
    \centering
    \small
    \includegraphics[trim={0cm 0cm 0cm 0cm},clip,width=\textwidth]{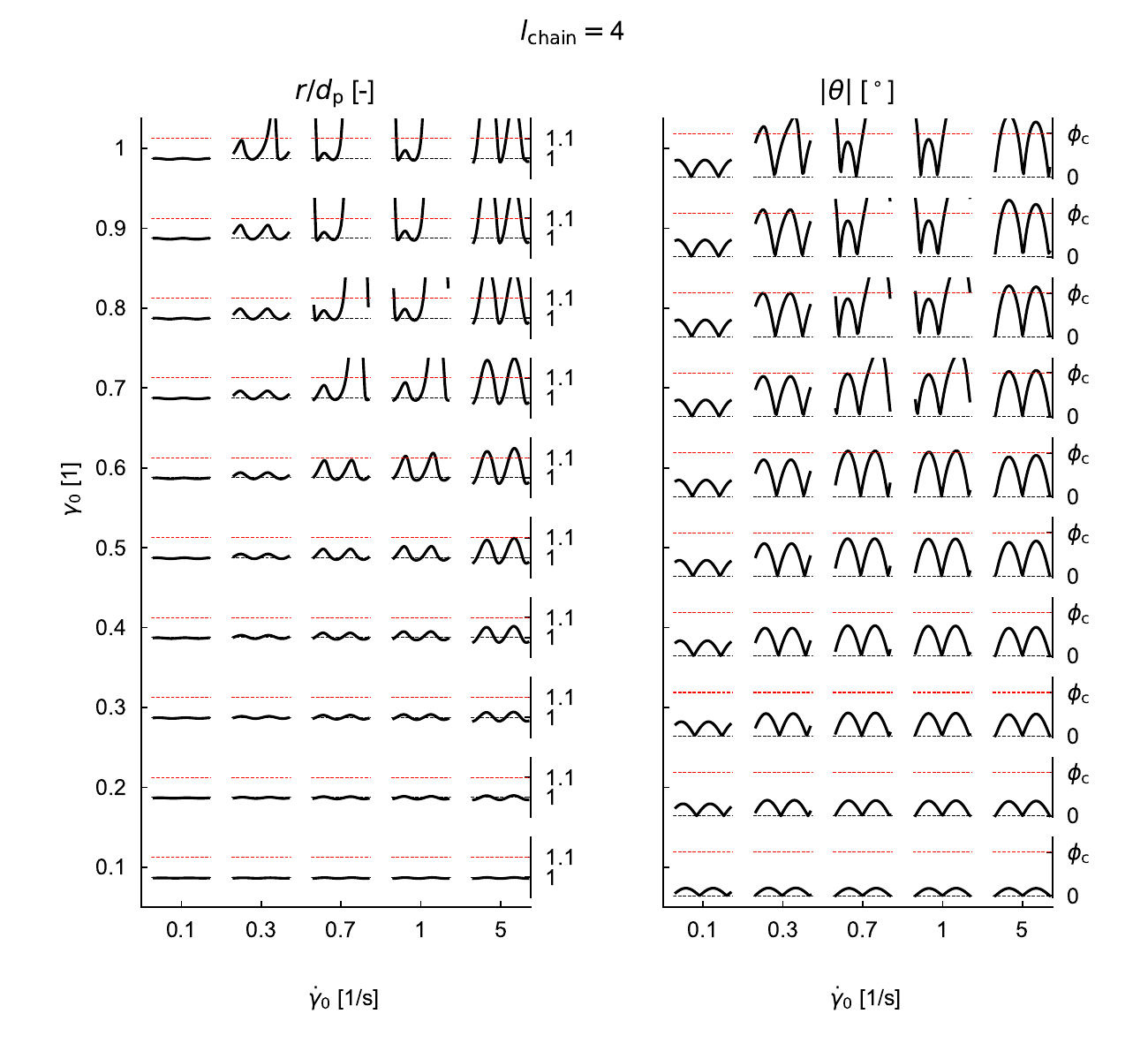}
    \caption{Behaviour of the central particle-pair of the chain with $l_\mathrm{chain}=4$ elements during the final simulated oscillatory shear cycle. (Left) scaled interparticle distance, $r/d_\mathrm{p}$, and (right) absolute value of chain angle, $\lvert\phi\rvert$. Chain break may be considered if $r/d_\mathrm{p}>1.1$ and $\lvert\phi\rvert>\phi_\mathrm{c}=33.5^\circ$.}
    \label{fig:Break_4}
\end{figure}

\begin{figure}[htp]
    \centering
    \small
    \includegraphics[trim={0cm 0cm 0cm 0cm},clip,width=\textwidth]{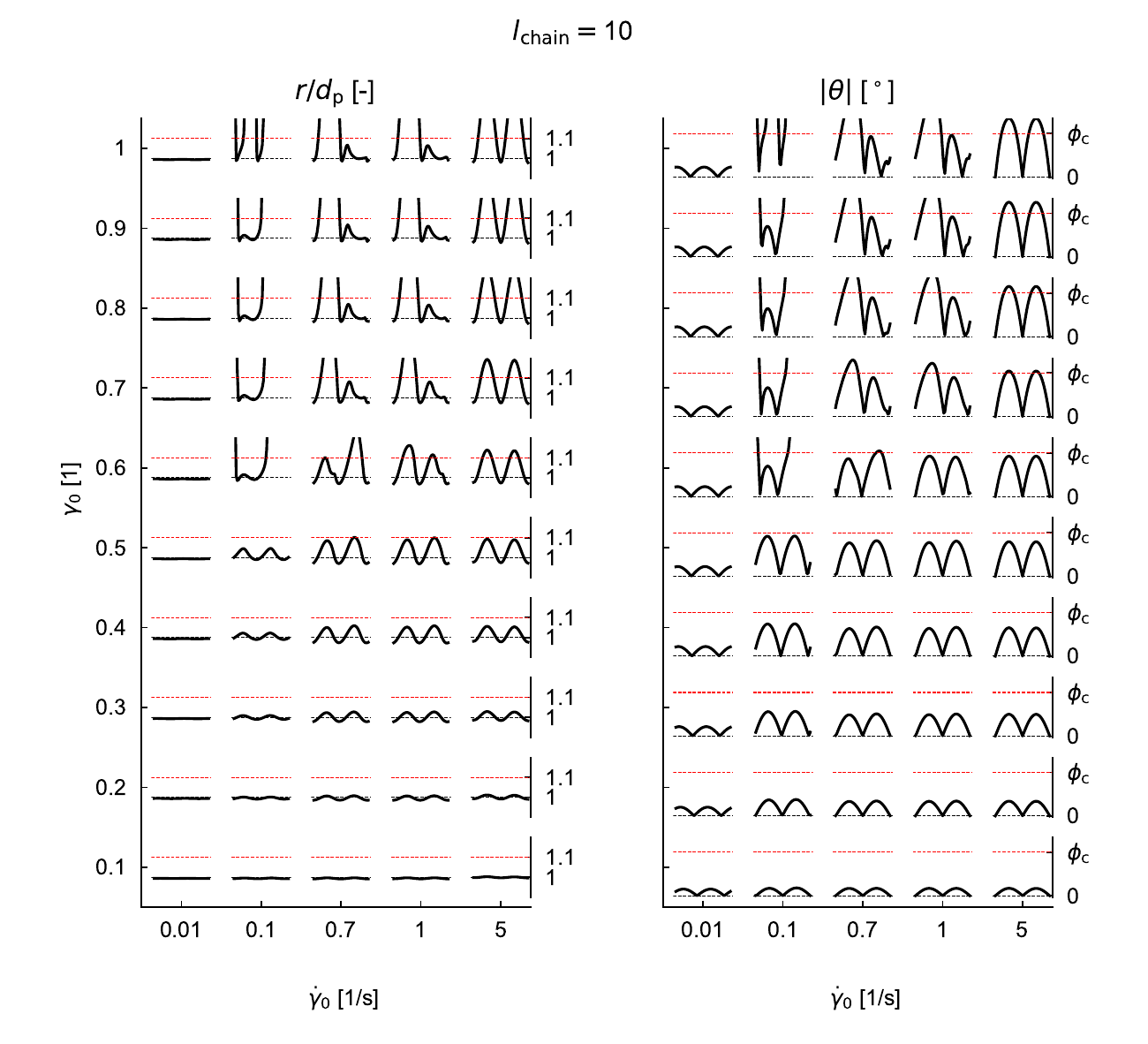}
    \caption{Behaviour of the central particle-pair of the chain with $l_\mathrm{chain}=10$ elements during the final simulated oscillatory shear cycle. (Left) scaled interparticle distance, $r/d_\mathrm{p}$, and (right) absolute value of chain angle, $\lvert\phi\rvert$. Chain break may be considered if $r/d_\mathrm{p}>1.1$ and $\lvert\phi\rvert>\phi_\mathrm{c}=33.5^\circ$.}
    \label{fig:Break_10}
\end{figure}

In Figures~\ref{fig:Break_4} and \ref{fig:Break_10} the instantaneous scaled interparticle distance, $r/d_\mathrm{p}$, and the absolute value of the chain angle, $\vert\phi\rvert$, of the central particle-pair on the final simulated cycle are shown, for $l_\mathrm{chain}=4$ and 10, respectively.

\pagebreak
\subsection{Magnetorheological alterations}
\label{sec:sup:comparisons}

To test the effects of the magnetic field density, particle concentration and particle type, these were independently varied from the benchmark measurement following the nomenclature previously presented in Table \ref{tab:working_fluids}. In the following Subsections are given, for each alteration, the raw Lissajous curves and first-harmonic moduli at different strains and frequencies for both analogues (NBa and VBa) as the continuous phase, and the Chebyshev coefficient spectrum (to the 15\textsuperscript{th} harmonic), the relative contribution of the third-order coefficients and relevant non-linear moduli ($G_1'$, $G_M'$, $G_L'$ and $\eta_1'$, $\eta_M'$, $\eta_L'$) for the NBa samples at the minimum tested oscillation frequency ($\omega=0.1$ rad/s).

The blue and red Lissajous curves represent, respectively, the elastic (stress vs strain) and viscous (stress vs strain rate) response. The first-harmonic moduli are plotted along with the experimental limits associated with low-torque issues and instrument and sample inertia (corrected for the geometry's gap-error).

\pagebreak
\subsubsection{15 wt\% MyOne @ 50 mT (15\_MyOne\_50)}
\begin{figure}[htp]
    \centering
    \begin{minipage}{\textwidth}
    \centering
    15\_MyOne\_50
    \end{minipage}
    \centering
    \begin{minipage}{\textwidth}
    \centering
    \small
    \includegraphics[trim={0cm 1cm 0cm 1.5cm},clip,width=\textwidth]{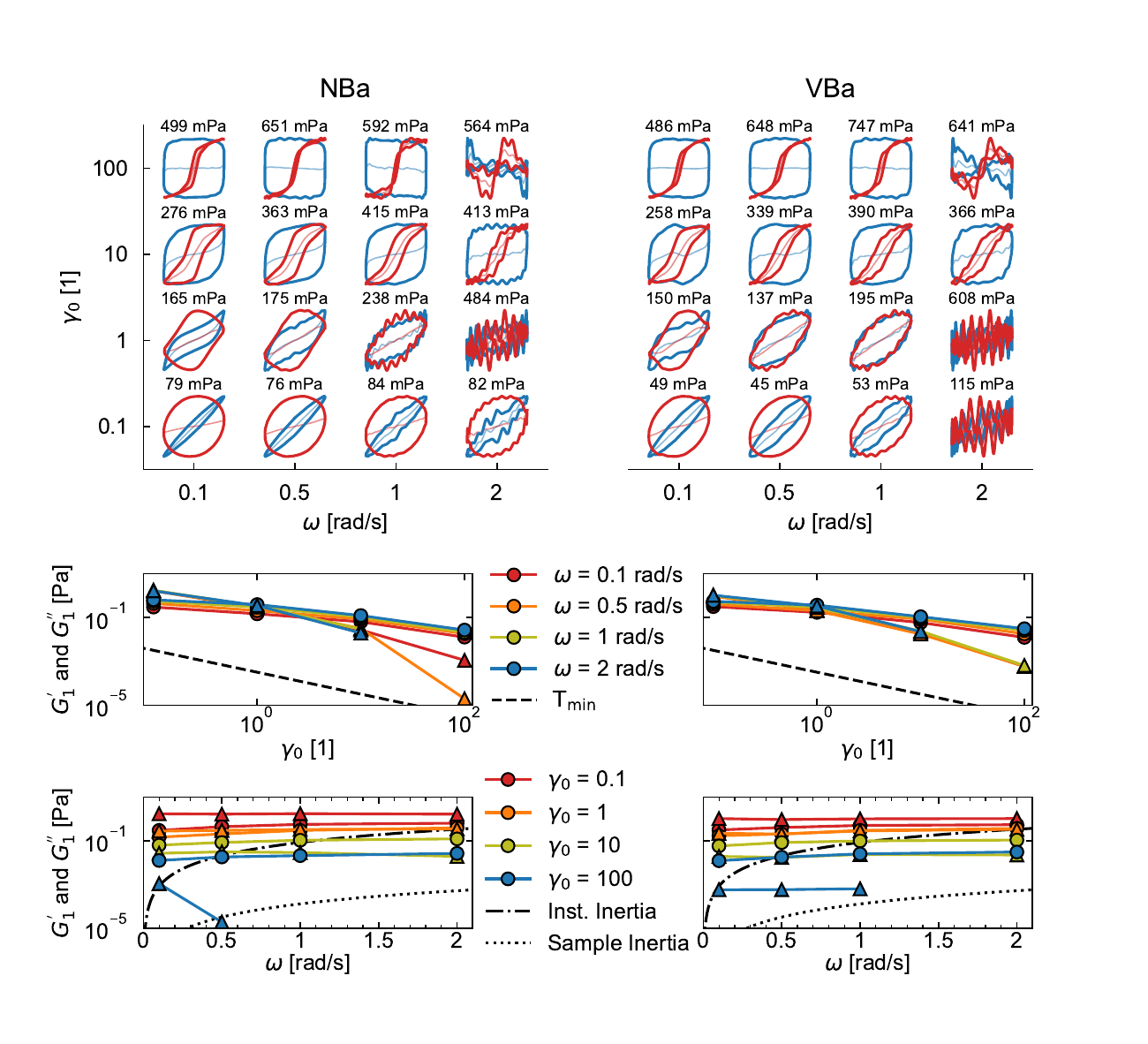}
    \end{minipage}
    \caption{Oscillatory shear data gathered with the (left) NBa and (right) VBa seeded with 15 wt\% of MyOne particles under a 50 mT magnetic field. (Top) Pipkin diagrams depicting the Lissajous curves and (bottom) First-harmonic loss ($G''_1$, in circular markers) and storage ($G'_1$, in triangular markers) moduli.}
    \label{fig:Oscillatory_15wt_MyOne_50mT}
\end{figure}
\begin{figure}[htp]
    \centering
    \begin{minipage}{\textwidth}
    \centering
    15\_MyOne\_50 ($\omega=0.1$ rad/s)
    \end{minipage}
    \centering
    \begin{minipage}{\textwidth}
    \centering
    \small
    \includegraphics[width=\textwidth]{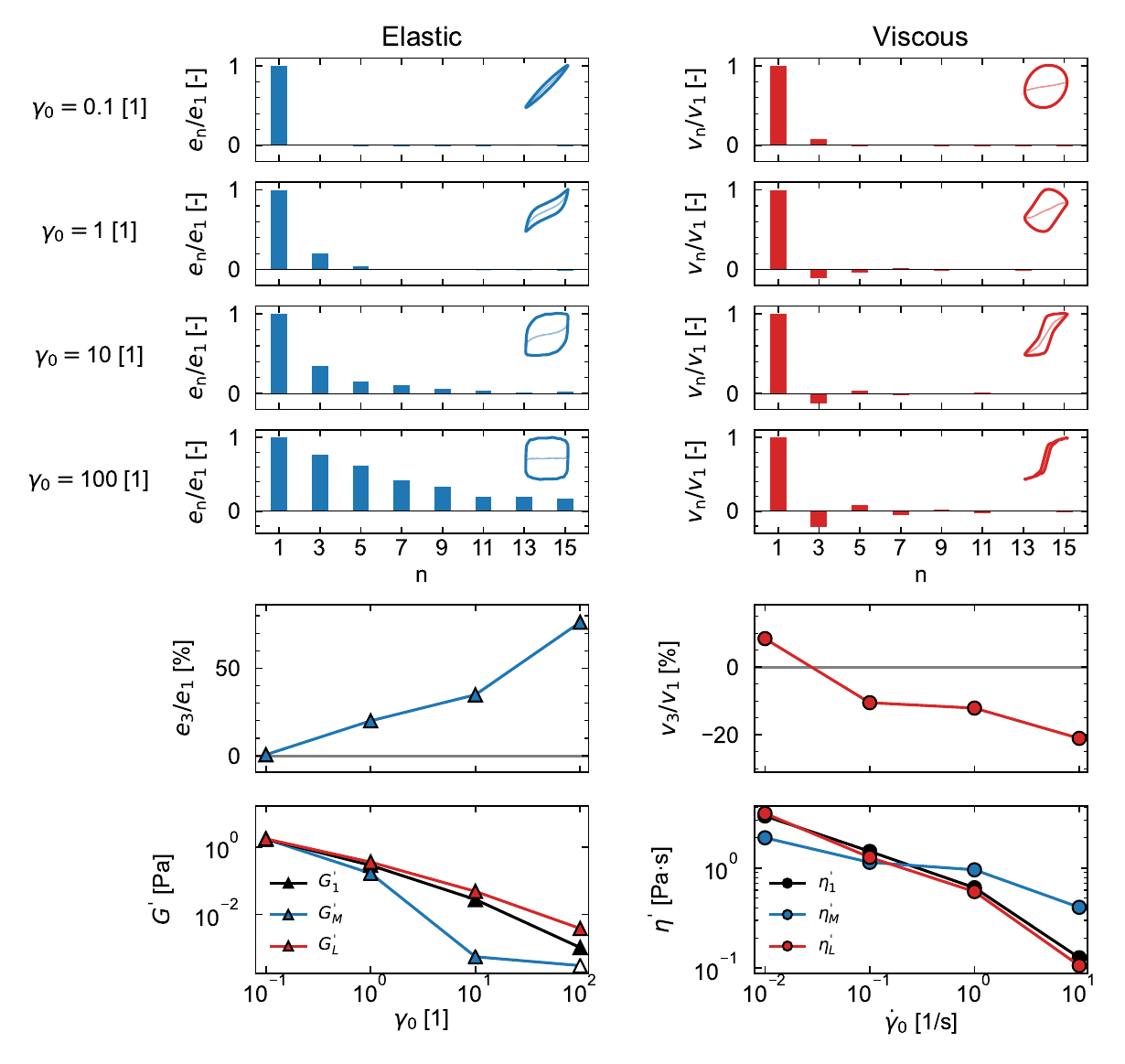}
    \end{minipage}
    \caption{Oscillatory shear data gathered with the NBa with 15 wt\% of MyOne particles under a 50 mT magnetic field, with the PP20 MRD P2 ($\omega=0.1$ rad/s). Scaled Chebyshev coefficient spectrum, $e_\mathrm{n}/e_1$ and $v_\mathrm{n}/v_1$ (with reconstructed Lissajous curves), variation of the scaled third harmonic coefficients, $e_3/e_1$ and $v_3/v_1$, and relevant viscoelasticity measures (elastic: $G_1'$, $G_M'$, $G_L'$, and viscous: $\eta_1'$, $\eta_M'$, $\eta_L'$) with applied deformation/deformation rate (negative $G_M'$ values are presented as open markers). Analysed with MITlaos\citep{mitlaos}.}
    \label{fig:15wt_MyOne_50mT_MITlaos}
\end{figure}

\pagebreak
\subsubsection{5 wt\% MyOne @ 250 mT (5\_MyOne\_250)}
\begin{figure}[htp]
    \centering
    \begin{minipage}{\textwidth}
    \centering
    5\_MyOne\_250
    \end{minipage}
    \centering
    \begin{minipage}{\textwidth}
    \centering
    \small
    \includegraphics[trim={0cm 1cm 0cm 1.5cm},clip,width=\textwidth]{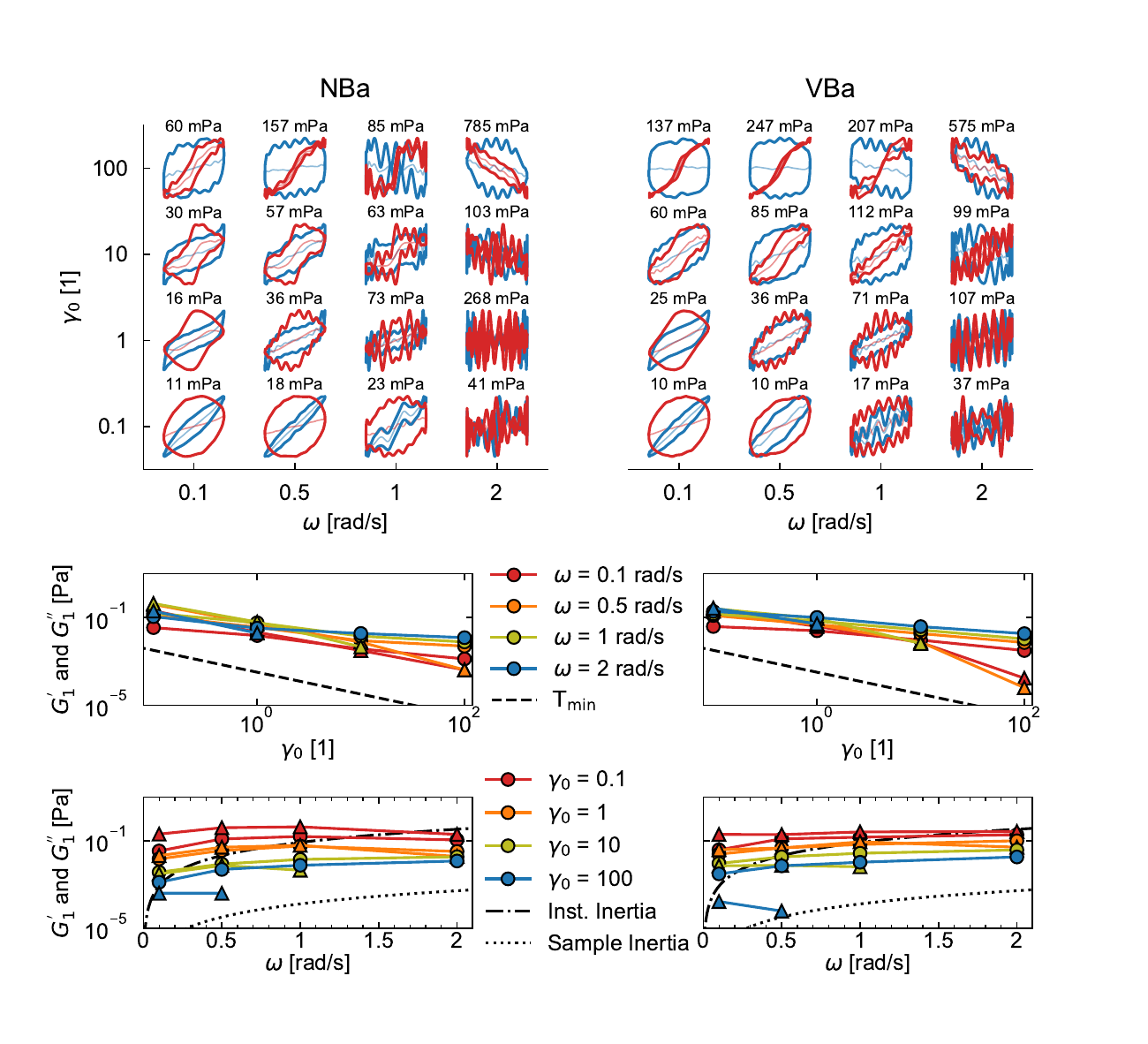}
    \end{minipage}
    \caption{Oscillatory shear data gathered with the (left) NBa and (right) VBa seeded with 5 wt\% of MyOne particles under a 250 mT magnetic field. (Top) Pipkin diagrams depicting the Lissajous curves and (bottom) First-harmonic loss ($G''_1$, in circular markers) and storage ($G'_1$, in triangular markers) moduli.}
    \label{fig:Oscillatory_5wt_MyOne_250mT}
\end{figure}
\begin{figure}[htp]
    \centering
    \begin{minipage}{\textwidth}
    \centering
    5\_MyOne\_250 ($\omega=0.1$ rad/s)
    \end{minipage}
    \centering
    \begin{minipage}{\textwidth}
    \centering
    \small
    \includegraphics[width=\textwidth]{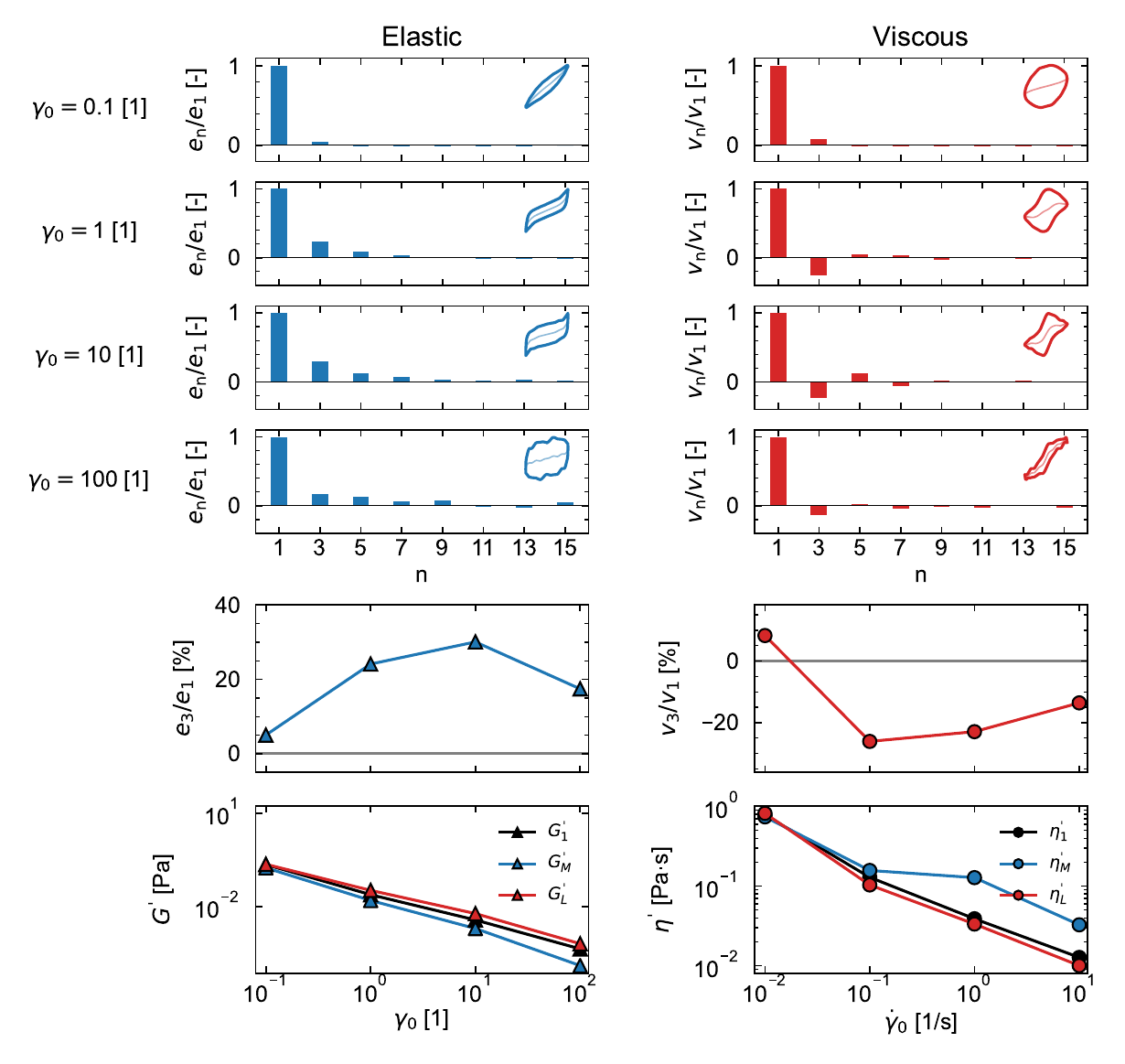}
    \end{minipage}
    \caption{Oscillatory shear data gathered with the NBa with 5 wt\% of MyOne particles under a 250 mT magnetic field, with the PP20 MRD P2 ($\omega=0.1$ rad/s). Scaled Chebyshev coefficient spectrum, $e_\mathrm{n}/e_1$ and $v_\mathrm{n}/v_1$ (with reconstructed Lissajous curves), variation of the scaled third harmonic coefficients, $e_3/e_1$ and $v_3/v_1$, and relevant viscoelasticity measures (elastic: $G_1'$, $G_M'$, $G_L'$, and viscous: $\eta_1'$, $\eta_M'$, $\eta_L'$) with applied deformation/deformation rate (negative $G_M'$ values are presented as open markers). Analysed with MITlaos\citep{mitlaos}.}
    \label{fig:5wt_MyOne_250mT_MITlaos}
\end{figure}

\pagebreak
\subsubsection{15 wt\% M270 @ 250 mT (15\_M270\_250)}
\begin{figure}[htp]
    \centering
    \begin{minipage}{\textwidth}
    \centering
    15\_M270\_250
    \end{minipage}
    \centering
    \begin{minipage}{\textwidth}
    \centering
    \small
    \includegraphics[trim={0cm 1cm 0cm 1.5cm},clip,width=\textwidth]{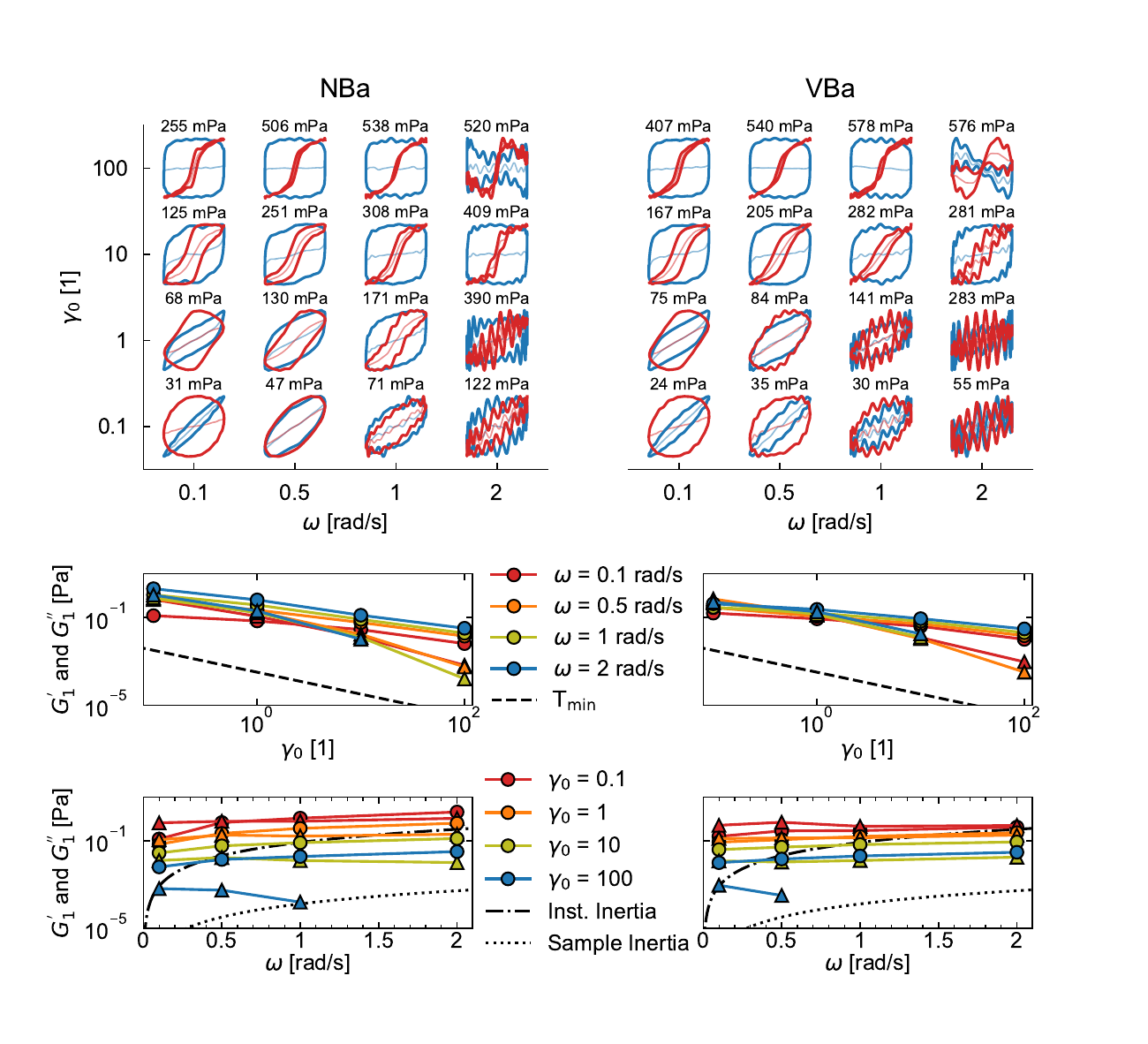}
    \end{minipage}
    \caption{Oscillatory shear data gathered with the (left) NBa and (right) VBa seeded with 15 wt\% of M270 particles under a 250 mT magnetic field. (Top) Pipkin diagrams depicting the Lissajous curves and (bottom) First-harmonic loss ($G''_1$, in circular markers) and storage ($G'_1$, in triangular markers) moduli.}
    \label{fig:Oscillatory_15wt_M270_250mT}
\end{figure}
\begin{figure}[htp]
    \centering
    \begin{minipage}{\textwidth}
    \centering
    15\_M270\_250 ($\omega=0.1$ rad/s)
    \end{minipage}
    \centering
    \begin{minipage}{\textwidth}
    \centering
    \small
    \includegraphics[width=\textwidth]{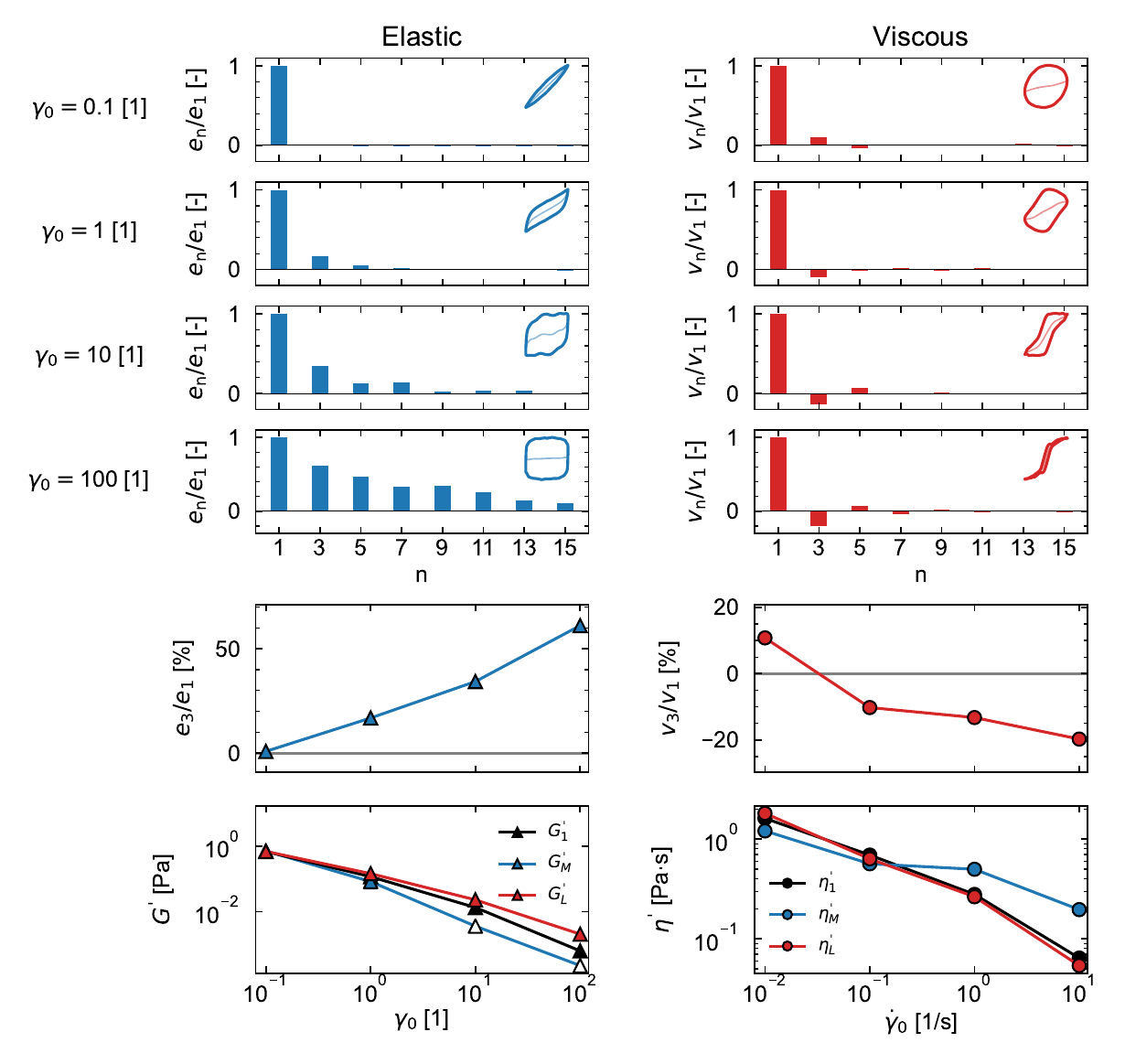}
    \end{minipage}
    \caption{Oscillatory shear data gathered with the NBa with 15 wt\% of M270 particles under a 250 mT magnetic field, with the PP20 MRD P2 ($\omega=0.1$ rad/s). Scaled Chebyshev coefficient spectrum, $e_\mathrm{n}/e_1$ and $v_\mathrm{n}/v_1$ (with reconstructed Lissajous curves), variation of the scaled third harmonic coefficients, $e_3/e_1$ and $v_3/v_1$, and relevant viscoelasticity measures (elastic: $G_1'$, $G_M'$, $G_L'$, and viscous: $\eta_1'$, $\eta_M'$, $\eta_L'$) with applied deformation/deformation rate (negative $G_M'$ values are presented as open markers). Analysed with MITlaos\citep{mitlaos}.}
    \label{fig:15wt_M270_250mT_MITlaos}
\end{figure}




\end{document}